\documentclass[12pt,letterpaper]{article}

\usepackage{amsmath}
\usepackage{amsfonts}
\usepackage{amssymb}
\usepackage{graphicx}
\usepackage{color}
\usepackage{booktabs}
\usepackage{inputenc}
\usepackage[T1]{fontenc}

\usepackage{graphicx}
\usepackage{epstopdf}
\usepackage{amsmath}
\usepackage{amsfonts}
\usepackage{amssymb}
\usepackage{color}
\usepackage{mathrsfs}
\usepackage{color}
\usepackage{enumerate}
\usepackage{xcolor,url}

\usepackage{cancel,dsfont} 

\newcommand\eea{\end{eqnarray}}
\newcommand\bea{\begin{eqnarray}}

\def\be{\begin{equation}}
\def\ee{\end{equation}}

\def\nn{\nonumber}

\newcommand{\dd}{\partial}
\def\dd{\partial}

\def\hinvMpc{h\,{\rm Mpc}^{-1}}
\def\lnAs{\ln (10^{10}A_s)}

\newcommand{\kmax}{k_{\rm max }}

\usepackage[colorlinks,bookmarks]{hyperref}
\definecolor{linkblue}{rgb}{0,0,0.8}
\definecolor{linkgreen}{rgb}{0,0.5,0}

\hypersetup{pdfpagemode=UseNone, pdfstartview=FitH, linkcolor=linkblue,
            citecolor=linkgreen, urlcolor=linkblue}

\newcommand{\Omegaref}{\vec{\Omega}_{\rm ref}}

\begin{document}

\begin{center}

{\Large \bf Efficient Cosmological Analysis of the SDSS/BOSS data\\[0.01cm]from\\[0.25cm] the Effective Field Theory of Large-Scale Structure}
\\[0.7cm]

{\large  Thomas Colas${}^{1,2,3}$, Guido d'Amico${}^{1,4}$,  Leonardo Senatore${}^{1,2}$,  \\[0.3cm]
Pierre Zhang${}^{5}$, Florian Beutler${}^{6}$\\[0.7cm]}

\end{center}

\begin{center}

\vspace{.6cm}

{\normalsize { \sl $^{1}$ Stanford Institute for Theoretical Physics, Physics Department,\\ Stanford University, Stanford, CA 94306}}\\
\vspace{.3cm}

{\normalsize { \sl $^{2}$ 
Kavli Institute for Particle Astrophysics and Cosmology,\\
 SLAC and Stanford University, Menlo Park, CA 94025}}\\
\vspace{.3cm}

{\normalsize { \sl $^{3}$ Departement de Physique, International Center of Fundamental Physics,\\
Ecole Normale Superieure,  24 rue Lhomond, 75231 Paris Cedex, France}}\\
\vspace{.3cm}

{\normalsize { \sl $^{4}$ Dipartimento di SMFI dell' Universita' di Parma e INFN Gruppo Collegato di Parma, Parma, Italy}}\\
\vspace{.3cm}

{\normalsize { \sl $^{5}$ Department of Astronomy, School of Physical Sciences, CAS Key Laboratory for Research in Galaxies and Cosmology, School of Astronomy and Space Science, \\
University of Science and Technology of China, Hefei, Anhui 230026, China}}\\
\vspace{.3cm}

{\normalsize { \sl $^{6}$ Institute of Cosmology \& Gravitation,  University of Portsmouth, Portsmouth, UK}}\\
\vspace{.3cm}

\vspace{.3cm}

\end{center}

\hrule \vspace{0.3cm}
{\small  \noindent \textbf{Abstract} \\[0.3cm]
\noindent  The precision of the cosmological data allows us to accurately approximate the predictions for cosmological observables by Taylor expanding up to a low order the dependence on the cosmological parameters around a reference cosmology. By applying this observation to the redshift-space one-loop galaxy power spectrum of the Effective Field Theory of Large-Scale Structure, we analyze the BOSS DR12 data by scanning over all the parameters of $\Lambda$CDM cosmology with massive neutrinos. We impose several sets of priors, the widest of which is just a Big Bang Nucleosynthesis prior on the current fractional energy density of baryons, $\Omega_b h^2$, and a bound on the sum of neutrino masses to be less than 0.9 eV. In this case we measure the primordial amplitude of the power spectrum, $A_s$, the abundance of matter, $\Omega_m$, the Hubble parameter, $H_0$, and the tilt of the primordial power spectrum, $n_s$, to about $19\%$, $5.7\%$, $2.2\%$ and $7.3\%$ respectively, obtaining $\lnAs=2.91\pm 0.19$, $\Omega_m=0.314\pm 0.018$, $H_0=68.7\pm 1.5$ km/(s Mpc) and $n_s=0.979\pm 0.071$ at 68\% confidence level. A public code is released with this preprint.

\vspace{0.3cm}}
\hrule

\vspace{0.3cm}
\newpage
\tableofcontents

\section{Introduction\label{sec:intro}}

In~\cite{DAmico:2019fhj}, the Effective Field Theory of Large-Scale Structure (EFTofLSS) has been applied to the analysis of the BOSS/SDSS data to extract the cosmological parameters. This represents an important milestone for this research approach, whose history we briefly highlight later in this introduction. In the analysis of~\cite{DAmico:2019fhj}, two parameters of the $\Lambda$CDM model, the tilt of the primordial power spectrum, $n_s$, and the ratio of the baryon versus dark matter abundance, $f_{bc}\equiv\Omega_b/\Omega_c$, were fixed to the preferred value of the Planck2018 data release~\cite{Aghanim:2018eyx}. Alternatively, when neutrinos were allowed to vary their mass, all parameters were fixed to the preferred value from Planck2018.  The purpose of this brief companion paper to~\cite{DAmico:2019fhj} is to eliminate these limitations.

In fact, there are two main reasons why some cosmological parameters of the analysis of~\cite{DAmico:2019fhj} have been fixed. First, on a more general level, the analysis of~\cite{DAmico:2019fhj} had the purpose of  originally establishing the usefulness of the EFTofLSS to analyze Large-Scale Structure data, and also to show how the same data could be powerfully used to measure some of the cosmological parameters in a way that is competitive with other probes. For example, within the aforementioned restrictions, Ref.~\cite{DAmico:2019fhj} measured $\Omega_m$ with error bars comparable to those of Planck2018~\cite{Aghanim:2018eyx}, and the present-day value of the Hubble rate, $H_0$, with error bars competitive with those from Supernovae surveys~\cite{Yuan:2019npk}. Given the size of the error bars on $f_{bc}$ and $n_s$ from Planck2018, it is clear that the general conclusions of~\cite{DAmico:2019fhj} are not significantly affected by the approximate treatment of these priors.

The second reason why the analysis of~\cite{DAmico:2019fhj} does not scan over the whole parameter space of $\nu\Lambda$CDM ({\it i.e.} $\Lambda$CDM plus massive neutrinos) is instead technical. The evaluation time of the prediction of the EFTofLSS for a given set of cosmological parameters is around 50 seconds on a normal CPU. This relative slowness is mainly due to the implementation of the IR-resummation, which is the original one developed in~\cite{Senatore:2014vja,Lewandowski:2018ywf}, that has the benefit of high accuracy but at the cost of a relatively long evaluation time. In order to efficiently evaluate the Likelihood of the cosmological parameters by running a Monte Carlo Markov Chain (MCMC), the authors of~\cite{DAmico:2019fhj} pre-computed the power spectra from the EFTofLSS on a grid. Size limitations forbid the creation of too large a grid.

In this companion paper, we eliminate this technical limitation of~\cite{DAmico:2019fhj} by exploiting the following fact. Cosmological data are by now precise enough to strongly limit the uncertainty range for all the cosmological parameters of interest, and for some of them the allowed range is extremely small. Since the cosmological predictions from the EFTofLSS (but this is true for any method of prediction) are smooth functions of the cosmological parameters,  it is expected, and indeed we will verify that this is true, that the whole range spanned by an observable as we scan over the allowed range of the cosmological parameters is  well approximated by a low-order  Taylor expansion of the EFTofLSS prediction around  a  reference  cosmology. This approach is not new  in the context of the EFTofLSS~\cite{Cataneo:2016suz}, but was never used beyond the prediction of the dark matter power  spectrum nor applied to data.

Ref.~\cite{DAmico:2019fhj}, and  this paper which should be understood as a companion, are the completion of a long journey for the EFTofLSS, which started with the initial development of the theory~\cite{Baumann:2010tm, Carrasco:2012cv,Porto:2013qua} and underwent many important steps in order for it to be compared to observational data. Before  starting, we mention a few of these results to provide a rough orientation to the interested reader.  The dark matter power spectrum has been computed at one-, two- and three-loop orders in~\cite{Carrasco:2012cv, Carrasco:2013sva, Carrasco:2013mua, Carroll:2013oxa, Senatore:2014via, Baldauf:2015zga, Foreman:2015lca, Baldauf:2015aha, Cataneo:2016suz, Lewandowski:2017kes,Konstandin:2019bay}. Some additional theoretical developments of the EFTofLSS that accompanied these calculations were a careful understanding of renormalization~\cite{Carrasco:2012cv,Pajer:2013jj,Abolhasani:2015mra} (including rather-subtle aspects such as lattice-running~\cite{Carrasco:2012cv} and a better understanding of the velocity field~\cite{Carrasco:2013sva,Mercolli:2013bsa}), of the several ways for extracting the value of the counterterms from simulations~\cite{Carrasco:2012cv,McQuinn:2015tva}, and of the non-locality in time of the EFTofLSS~\cite{Carrasco:2013sva, Carroll:2013oxa,Senatore:2014eva}. These theoretical explorations also include an instructive study in 1+1 dimensions~\cite{McQuinn:2015tva}. In order to correctly describe the Baryon Acoustic Oscillation~(BAO) peak, an IR-resummation of the long displacement fields had to be performed. This has led to the so-called IR-resummed EFTofLSS~\cite{Senatore:2014vja,Baldauf:2015xfa,Senatore:2017pbn,Lewandowski:2018ywf,Blas:2016sfa}.  A method to account for baryonic effects was presented in~\cite{Lewandowski:2014rca}. The dark-matter bispectrum has been computed at one-loop in~\cite{Angulo:2014tfa, Baldauf:2014qfa}, the one-loop trispectrum in~\cite{Bertolini:2016bmt}, 
the displacement field in~\cite{Baldauf:2015tla}. The lensing power spectrum has been computed at two loops in~\cite{Foreman:2015uva}.  Biased tracers, such as halos and galaxies, have been studied in the context of the EFTofLSS in~\cite{ Senatore:2014eva, Mirbabayi:2014zca, Angulo:2015eqa, Fujita:2016dne, Perko:2016puo, Nadler:2017qto} (see also~\cite{McDonald:2009dh}), the halo and matter power spectra and bispectra (including all cross correlations) in~\cite{Senatore:2014eva, Angulo:2015eqa}. Redshift space distortions have been developed in~\cite{Senatore:2014vja, Lewandowski:2015ziq,Perko:2016puo}. Clustering dark energy has been included in the formalism in~\cite{Lewandowski:2016yce,Lewandowski:2017kes,Cusin:2017wjg,Bose:2018orj}, primordial non-Gaussianities in~\cite{Angulo:2015eqa, Assassi:2015jqa, Assassi:2015fma, Bertolini:2015fya, Lewandowski:2015ziq, Bertolini:2016hxg}, and neutrinos in~\cite{Senatore:2017hyk,deBelsunce:2018xtd}. Faster evaluation schemes for evaluation for some of the loop integrals have been developed in~\cite{Simonovic:2017mhp}.

The EFTofLSS became ready to be compared with observational data of galaxy clustering with the completion of~\cite{Perko:2016puo}, as only at that point the IR-resummed one-loop power spectrum of biased tracers in redshift space in $\Lambda$CDM cosmology had been computed. This is where the journey of Ref.~\cite{DAmico:2019fhj} and of this paper begins. Since this paper represents, as mentioned, a small but important extension of the analysis pipeline of~\cite{DAmico:2019fhj}, we refer the reader to~\cite{DAmico:2019fhj} for all the introductory and overview material about the EFTofLSS, as well as for some physical explanations, and for  some comparisons of the performance of the EFTofLSS in terms of measurement of cosmological parameters against numerical simulations. Here we focus only on the non-trivial results that are novel in this paper.\\

With this publication, we release a C++ code that computes the IR-resummed one-loop power spectrum multipoles and tree-level bispectrum monopole of biased tracers in the EFTofLSS, called CBiRd, and includes a Python-based construction of the approximation by Taylor expansion: Code for Biased tracers in Redshift space at 
\href{https://github.com/pierrexyz/cbird}{CBiRd GitHub} 
(see also for a repository of all EFTofLSS codes: \href{http://stanford.edu/~senatore/}{EFTofLSS repository}).\\

{\bf Note Added:} Ref.~\cite{Ivanov:2019pdj}, which has just appeared, has some significant overlap with this work.

\section{Taylor-Expanded Functional Dependence}

The cosmological parameters affect the predictions of the EFTofLSS in two main ways: through the linear power spectrum, and through the EFT-coefficients that encode the effect of short-distance non-linearities  at long distances (for example the so-called speed of  sound or bias coefficients). Exploiting the cosmology dependence of the EFT parameters requires to be able to predict the physics within the non-linear regime as a function of the different cosmologies. Though this is a promising avenue to follow (as, for example, it would improve the constraining power of the EFTofLSS), in this paper, as in~\cite{DAmico:2019fhj}, we decide to be completely agnostic about such a dependence.

Instead, clearly, the cosmological dependence of the linear and one-loop parts of the power spectrum, or of the IR-resummation cannot be neglected. However, since observational data force us to explore a limited range of cosmological parameter space, where the observables are allowed to change very little around a reference cosmology, we can Taylor expand the EFT prediction around this reference cosmology to a low order in the deviation of the cosmological parameters from the reference cosmology. In formulas, if $\vec \Omega$ represents the set of cosmological parameters, with $\Omegaref$ the reference cosmology, and $\vec b$ the EFT-parameters, we can write the EFT prediction for any observable in the following way. Here for simplicity we will focus on the power spectrum, and we take the model to be the same as in~\cite{DAmico:2019fhj}, to which we refer the reader for a detailed description. If $P_{\rm EFT}(k,\vec b, \vec \Omega)$ is the EFT prediction, we can write:
\bea
&&P_{\rm EFT}(k,\vec b, \vec \Omega)=P_{\rm EFT}(k,\vec b,  \Omegaref)+\left. \frac{\dd}{\dd\Omega_i}P_{\rm EFT}(k,\vec b, \vec \Omega)\right|_{\Omegaref} \left(\vec \Omega-\ \Omegaref\right)_i+\\ \nn
&&\qquad\qquad+\frac{1}{2}\left.\frac{\dd^2}{\dd\Omega_i\dd\Omega_j}P_{\rm EFT}(k,\vec b, \vec \Omega)\right|_{\Omegaref} \left(\vec \Omega-\ \Omegaref\right)_i\left(\vec \Omega-\ \Omegaref\right)_j+\ldots \ ,
\eea
where $\ldots$ represent higher order terms in $\left(\vec \Omega-\Omegaref\right)$. We can then decide the order of the Taylor expansion and then evaluate the required derivatives with some finite difference method to the desired accuracy. 

In this paper, we focus on the analysis of the power spectra monopole and quadrupole of the BOSS D12 data (see for example~\cite{Reid:2015gra, Alam:2016hwk,Beutler:2016arn}).  As we will show, to perform such an analysis it is enough to expand $P_{\rm EFT}(k,\vec b, \vec \Omega)$ just to third order in all the cosmological parameters. 
We summarize the specifications of the Taylor expansion we use for this analysis in Table~\ref{tab:Taylorderivative}.  We however stress that our code is already set up to include up to the fifth derivative. Notice that the Taylor expansion is performed directly at the level of the observable quantity that is being compared to data, that is after the application of the Alcock-Paczynski effect~\cite{Alcock:1979mp}, of the window function and of the fiber collision correction~\cite{Hahn:2016kiy}, all of which are applied as in~\cite{DAmico:2019fhj}.

The cosmological parameters we choose for the Taylor expansion are $A_s$, $h$, $\omega_c$, $\omega_b$, $n_s$, $\sum_i m_{\nu_i}$, where $h$ is the present value of the Hubble constant in units of 100\,km/(s Mpc), $ \sum_i m_{\nu_i}$ is the sum of the neutrino masses, that for simplicity we take to follow the normal hierarchy (NH), and $\omega_b=\Omega_b h^2$, with $\Omega_b$ being the baryon abundance. {Since we evaluate the derivatives with a finite difference method with second order accuracy, we need a minimum number of $4^n$ evaluations of the power spectra, where $n$ is the number of parameters. To this we add the reference cosmology evaluation.} Such a small number of evaluations does not seem to represent a numerical challenge~\footnote{\label{footnote:lower_conv}Furthermore, we did not make any significant effort in optimization.  For example, one could consider a different number of derivatives for each cosmological parameter. For instance, by fixing the neutrino masses, we found that already truncating the Taylor  expansion to linear order would have  actually been sufficient for the analysis of the CMASS NGC sample of the BOSS data, {while a second order expansion is enough to provide constraints and most-likely values not farther than $ \sigma_{\rm stat}/5$ from the ones obtained with the third order expansion, which we use for all analyses we performed here.} }.

\begin{table}
\centering
\footnotesize
\begin{tabular}{|l|l|l|l|l|l|l|} 
\hline
Cosmological Parameter          & Order of Derivative & Size of Step & Reference value  \\ \hline
$A_s$                 & 3    & 10 \%     &  $1.7116 \times 10^{-9}$     \\ \hline
$h$     & 3                           & 2 \%     &    0.690    \\ \hline
$\omega_c$ & 3                  & 4 \%   & 0.1285            \\ \hline
$\omega_b$     & 3           & 4 \%     & 0.02339     \\ \hline
$n_s$     & 3                           & 2 \%     &  0.9649      \\ \hline
$\sum_i m_{\nu_i}$ [eV] (NH)    & 3                           & 25 \%      &   0.4     \\ \hline
\end{tabular}
     \caption{\small Technical specifications of the Taylor expansion for the data analysis. The size of the step is expressed as percentage of reference value.}
     \label{tab:Taylorderivative}
\end{table}

\begin{figure}[h!]
\centering
\includegraphics[width=0.47\textwidth,draft=false]{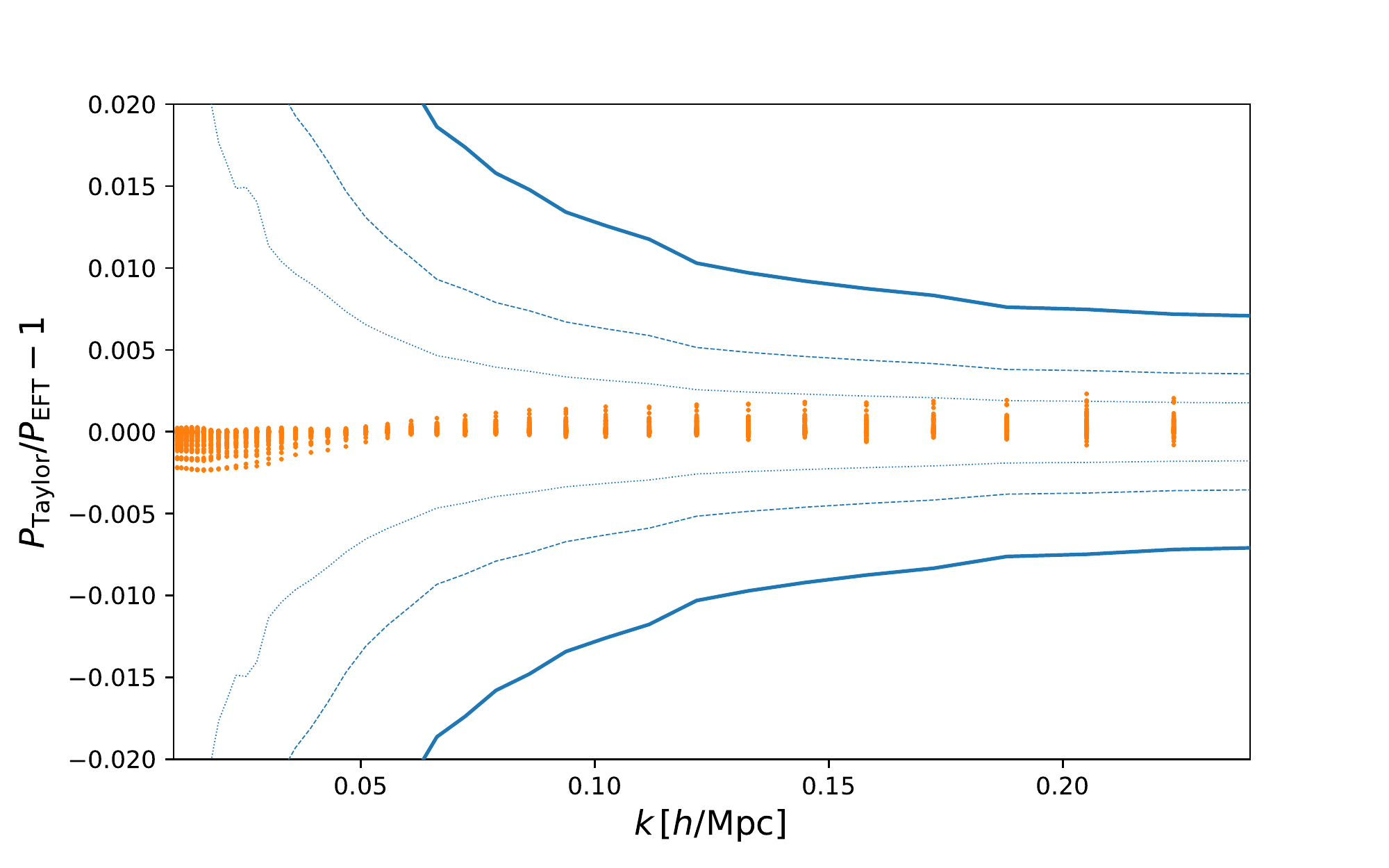}
\includegraphics[width=0.47\textwidth,draft=false]{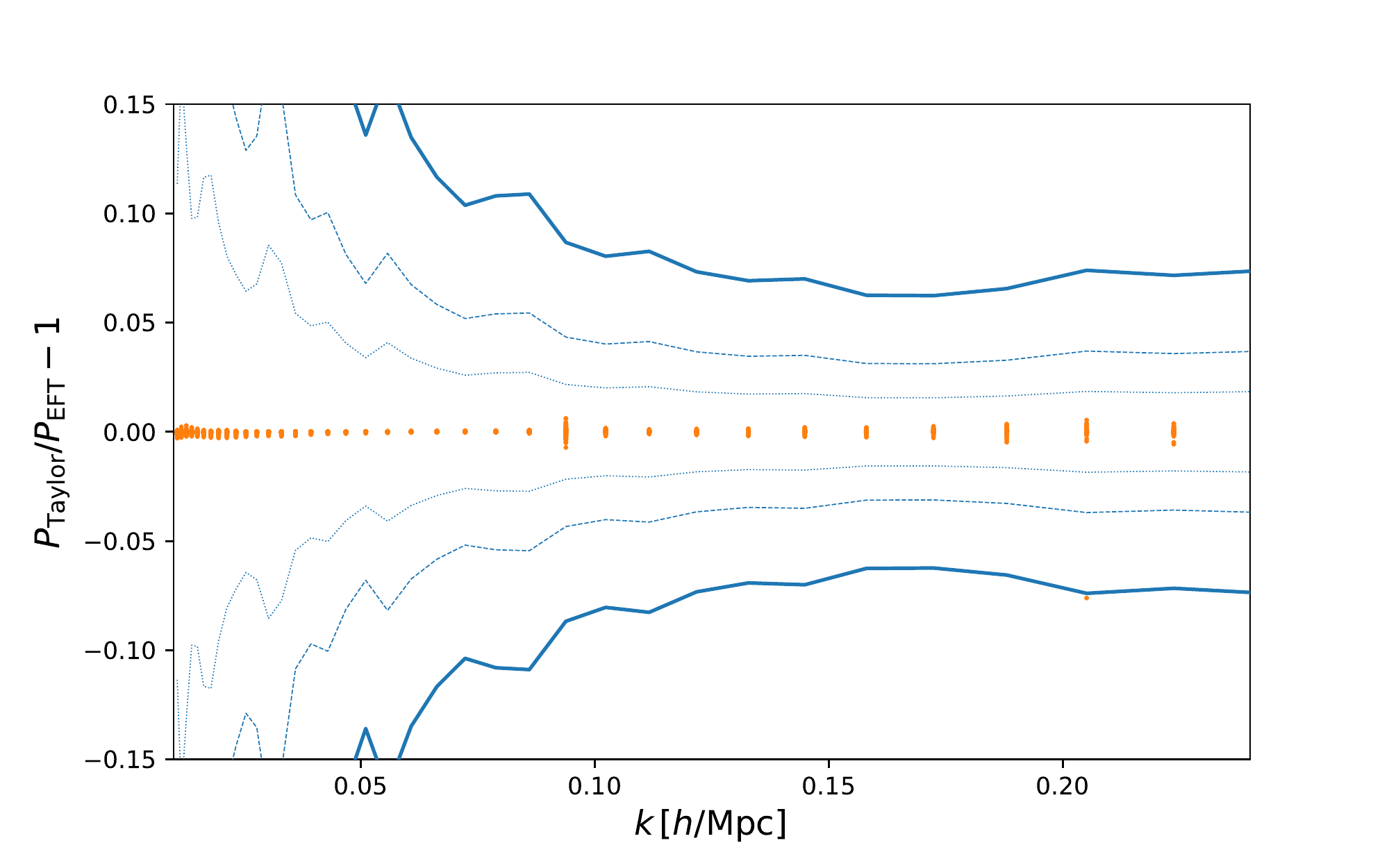}
\includegraphics[width=0.47\textwidth,draft=false]{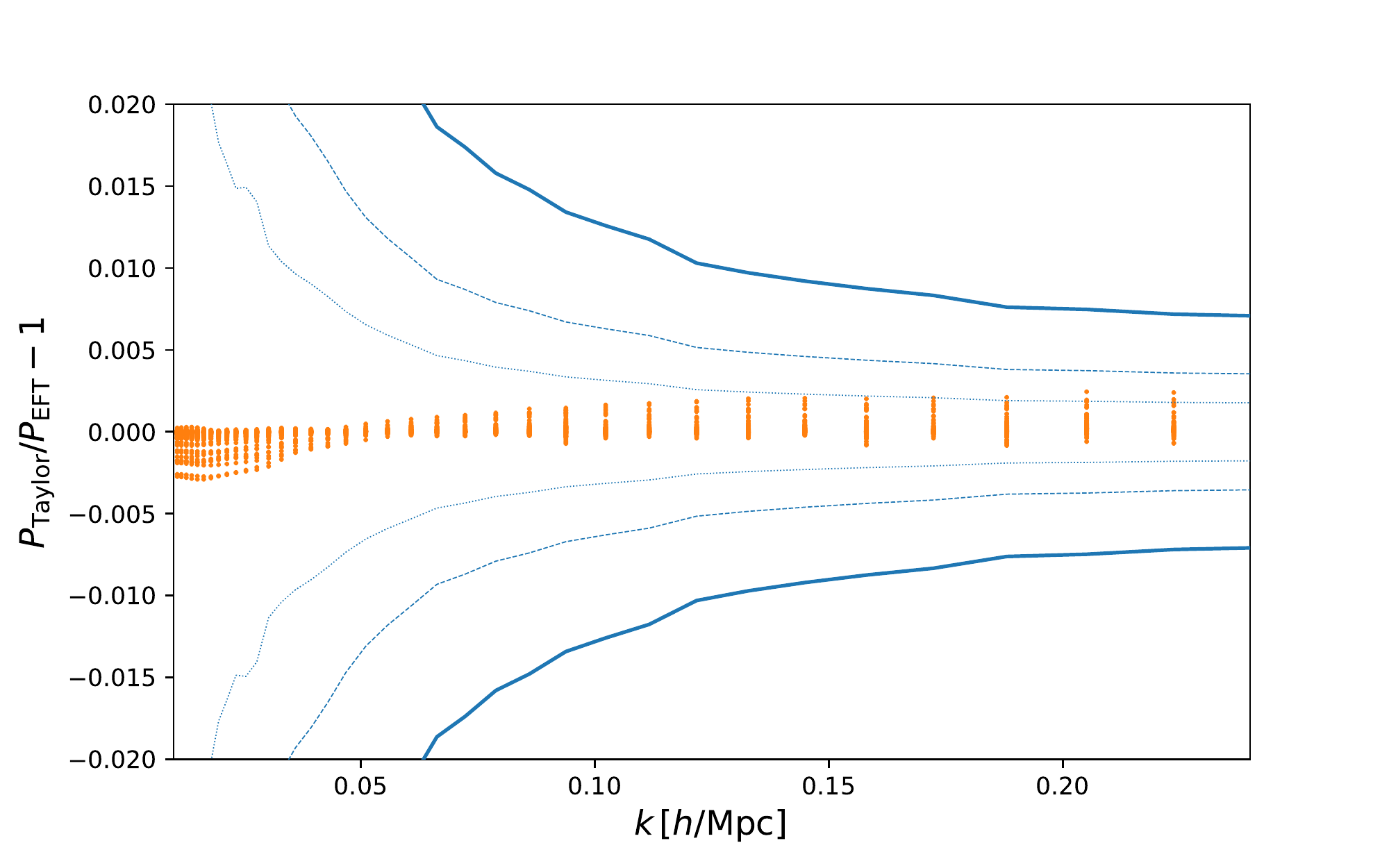}
\includegraphics[width=0.47\textwidth,draft=false]{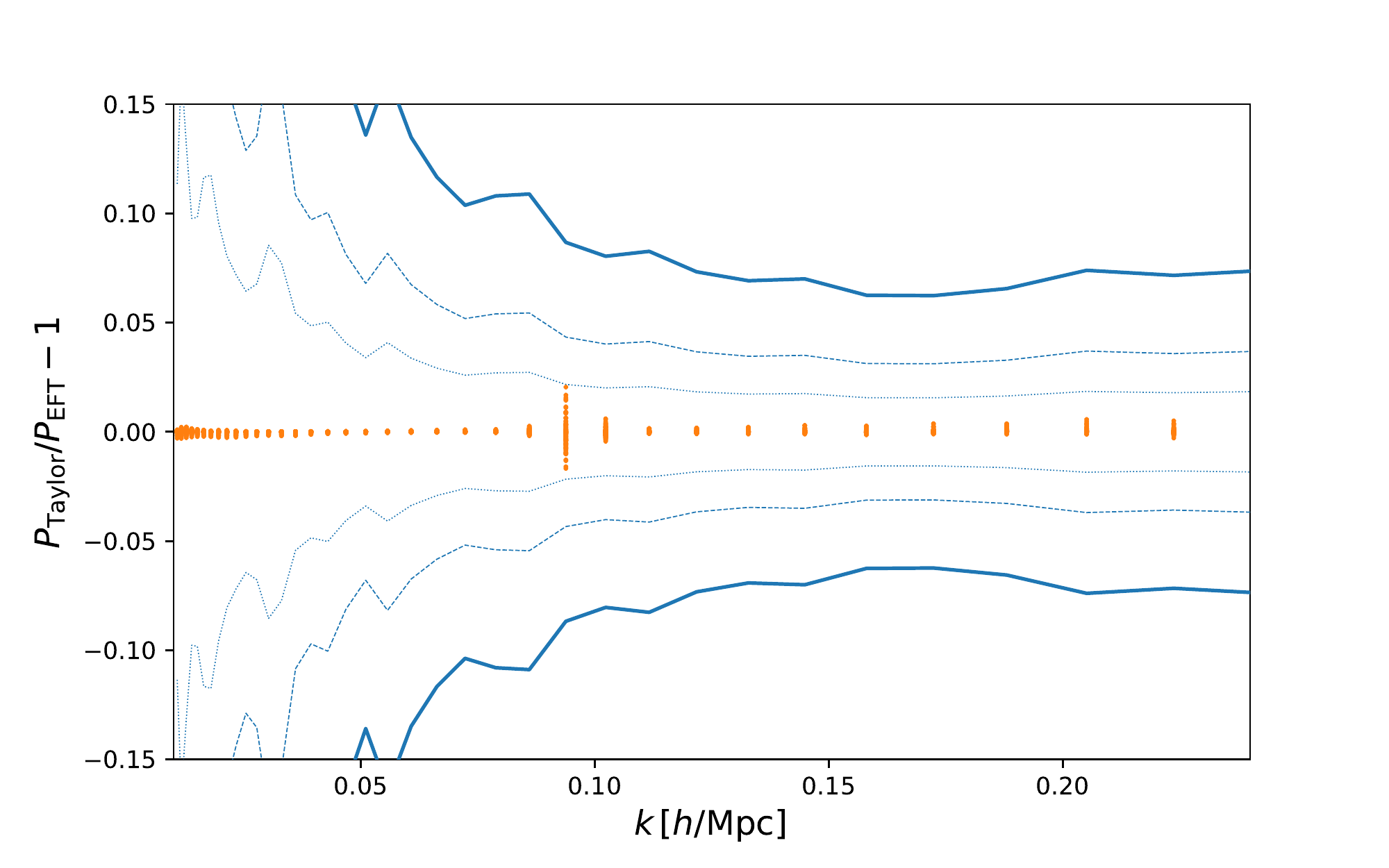}
\caption{\small {\it Top}: Relative difference  between the computation of an EFTofLSS power spectrum by the direct evaluation or by approximation with the Taylor expansion, 
for cosmologies that lie on a hyper-ellipsoid with orthogonal semi-axes of length  $2 \sigma_{\rm stat}$ centered on the reference cosmology, where the $2\sigma_{\rm stat}$'s are, for each cosmological parameter, the largest 95\% confidence intervals obtained amongst the analyses with various priors performed here. Explicitly, we choose the following $2\sigma_{\rm stat}$ for the parameters: $30\%$ for $A_s$, $9.4\%$ for~$\Omega_m$, $8.4\%$ for~$h$, $12\%$ for $n_s$, for $\omega_b$, we choose  $5.3\%$, which is motivated by BBN. For the neutrinos, we take $\sum_i m_{\nu_i} \in [0.1, 0.8]$eV. 
On the left we plot the monopole, on the right the quadrupole. 
In solid blue, we plot the $1\sigma$ error bars of the CMASS data, and in dashed and dotted blue the $\sigma/2$ and $\sigma/4$ error bars of the data, respectively. {\it Bottom:} Same as the top figure, but with the $2\sigma_{\rm stat}$ interval for $\omega_b$ equal to 20\%.
In both plots, we see that the disagreement is very small when confronted to the error bars of the data, and indeed it has negligible consequences on the inferred cosmological parameters.
}
\label{fig:relativepower}
\end{figure}

We perform the following two tests to validate the accuracy of the Taylor expansion. In the first test,  in Fig.~\ref{fig:relativepower}, we compare the relative difference between the computation of a EFTofLSS power spectrum by the direct evaluation or by approximation with the Taylor expansion, for cosmologies that lie on an 
hyper-ellipsoid with orthogonal semi-axes of length  $2 \sigma_{\rm stat}$ centered on the reference cosmology, where $\sigma_{\rm stat}$ are the largest error bars we will obtain for each parameter from the various analyses presented here. We use the $\sigma_{\rm stat}$'s obtained from the analysis of CMASS sample with the BBN priors, and at the same time we keep the larger error bar on $h$ from the analysis with prior of~$f_{bc}$. Explicitly, we choose the following $2\sigma_{\rm stat}$ for the parameters: $30\%$ for $A_s$, $9.4\%$ for~$\Omega_m$, $8.4\%$ for~$h$, $12\%$ for $n_s$, for $\omega_b$, we choose  $5.3\%$~({\it top}) and  $20\%$~({\it bottom}). The $2\sigma_{\rm stat}=5.3\%$ on $\omega_b$ is motivated by Big Bang Nucleosynthesis (BBN). For neutrino masses, given that we have no measurements, we choose the overall interval to be $\sum_i m_{\nu_i} \in [0.1, 0.8]$eV. To further clarify the accuracy of the Taylor expansion, in App.~\ref{app:furtherchecks}, we provide a plot similar to Fig.~\ref{fig:relativepower}, but using $3\sigma_{\rm stat}$ instead of $2\sigma_{\rm stat}$, albeit with the slightly smaller $\sigma_{\rm stat}$ associated to the sample CMASSxLOWZ.  In the same Appendix, we show some relative differences between direct evaluation and Taylor expansion as we change one parameter for ranges significantly much larger than the ones that are probed here, in order to better investigate the overall accuracy of the Taylor expansion. The results of App.~\ref{app:furtherchecks} further confirm the accuracy of the Taylor expansion for the analyses we are interested in.

\begin{figure}[h!]
\centering
\includegraphics[width=0.495\textwidth,draft=false]{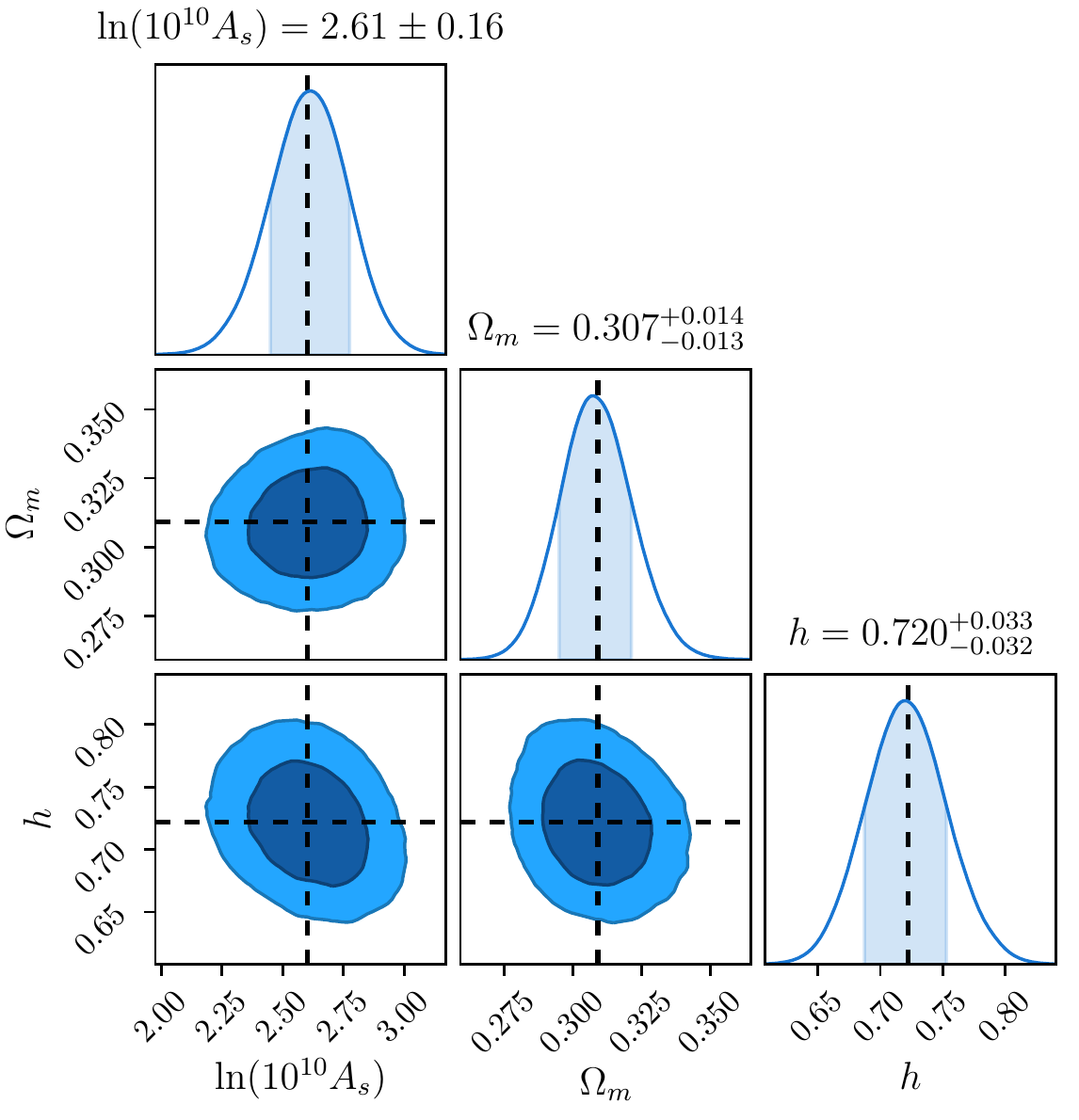}
\includegraphics[width=0.495\textwidth,draft=false]{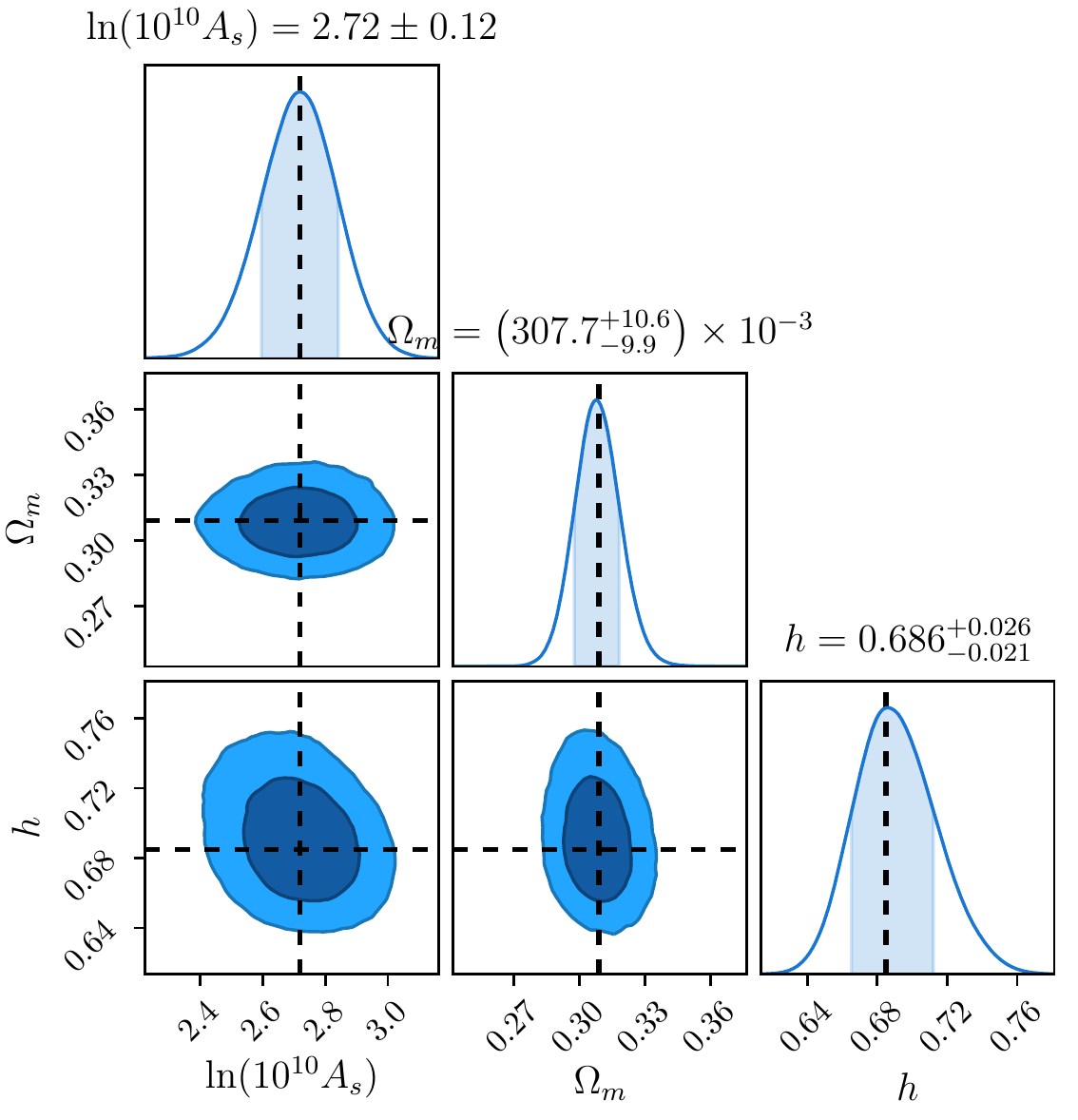}
\caption{\small {\it Left:} Posterior distributions of $\lnAs$, $\Omega_m$ and $h$ obtained from the analysis of the CMASS NGC sample when fixing $f_{bc},\, n_s$ and $\sum_i m_{\nu_i}$ as in~\cite{DAmico:2019fhj}, up to $\kmax=0.20\hinvMpc$ using the Taylor expansion approximation for the EFT power spectrum. This figure should reproduce exactly Fig. 14~\cite{DAmico:2019fhj}, where the same data were  analyzed using a grid (see also Table 4 there). In vertical dashed we plot the expectation value from~\cite{DAmico:2019fhj}.  We find that the disagreement is negligible for all 3 parameters, showing the great accuracy of the Taylor expansion for this dataset. {\it Right:} Same but for the CMASS sample combined with LOWZ NGC sample up to $\kmax=0.20\hinvMpc$ for CMASS and up to $\kmax=0.18\hinvMpc$ for LOWZ NGC. Now one should compare with Fig. 17~\cite{DAmico:2019fhj}. The agreement is again remarkably good, though one should keep in mind that here we used $\sum_i m_{\nu_i} = 0.06$eV with NH instead of a single massive neutrino.
}
\label{fig:NGChistogram}
\end{figure}

The second test that we  perform to check the  accuracy of the Taylor expansion is to analyze the  North  Galactic Cap (NGC) of the CMASS sample and also the whole CMASS sample combined with LOWZ NGC sample by fixing $n_s$ and  $\Omega_b/\Omega_c$ to the Planck2018 preferred values, and compare with the results of~\cite{DAmico:2019fhj}. For CMASS NGC analysis, we choose $\Omegaref$ to be close to the peak of the distribution, while for the test on CMASS combined with LOWZ NGC we run using the Taylor expansion used for all the other analysis, specified in Table~\ref{tab:Taylorderivative}. The results are plotted in Fig.~\ref{fig:NGChistogram}, where we see that the difference of the results is negligibly small ({\it i.e.} almost within the truncation error). This test does not check for the accuracy of the Taylor expansion in the direction of varying $n_s,\;\Omega_b/\Omega_c$ and, at some level, $\sum_{i}m_{\nu_i}$. However, the priors we will impose on them are so stringent or the interval we will explore so limited so that, together with the results of Fig.~\ref{fig:relativepower} and of App.~\ref{app:furtherchecks}, we feel assured that the dependence on these parameters, over the range we are  going to consider, is very well reproduced.
Furthermore, for $\sum_{i}m_{\nu_i}$, notice that we are running the analysis on CMASS combined with LOWZ NGC with the Taylor expansion centered at $0.4$eV, and imposing the sum to be $0.06$eV, and comparing with an MCMC run with a grid evaluated  with $0.06$eV total mass: therefore this test actually checks that the error of the Taylor expansion from exploring far away neutrino masses is negligible in this regime.

We end this section with the  following observation. The former results check the accuracy of the Taylor expansion only for a large but finite range of departure from the reference cosmology. Of course, if the posterior distribution of some cosmological parameter is peaked very far from the  reference cosmology, the accuracy of the  Taylor  expansion  at some point might become too low. We have found that the results for the cosmological parameters are accurate even when the peak of the distribution are  further than what is represented in Fig.~\ref{fig:relativepower}. However we point out that the construction of the Taylor expansion and the analysis through the MCMC is so fast, that one can iterate  the construction of the Taylor expansion by adjusting the parameters of the reference cosmology, or alternatively carry the Taylor expansion to an higher order.

\section{Cosmological Analysis of the BOSS data}

\subsection{Minimal-Mass Neutrinos}

\begin{figure}[h!]
\centering
\includegraphics[width=0.495\textwidth,draft=false]{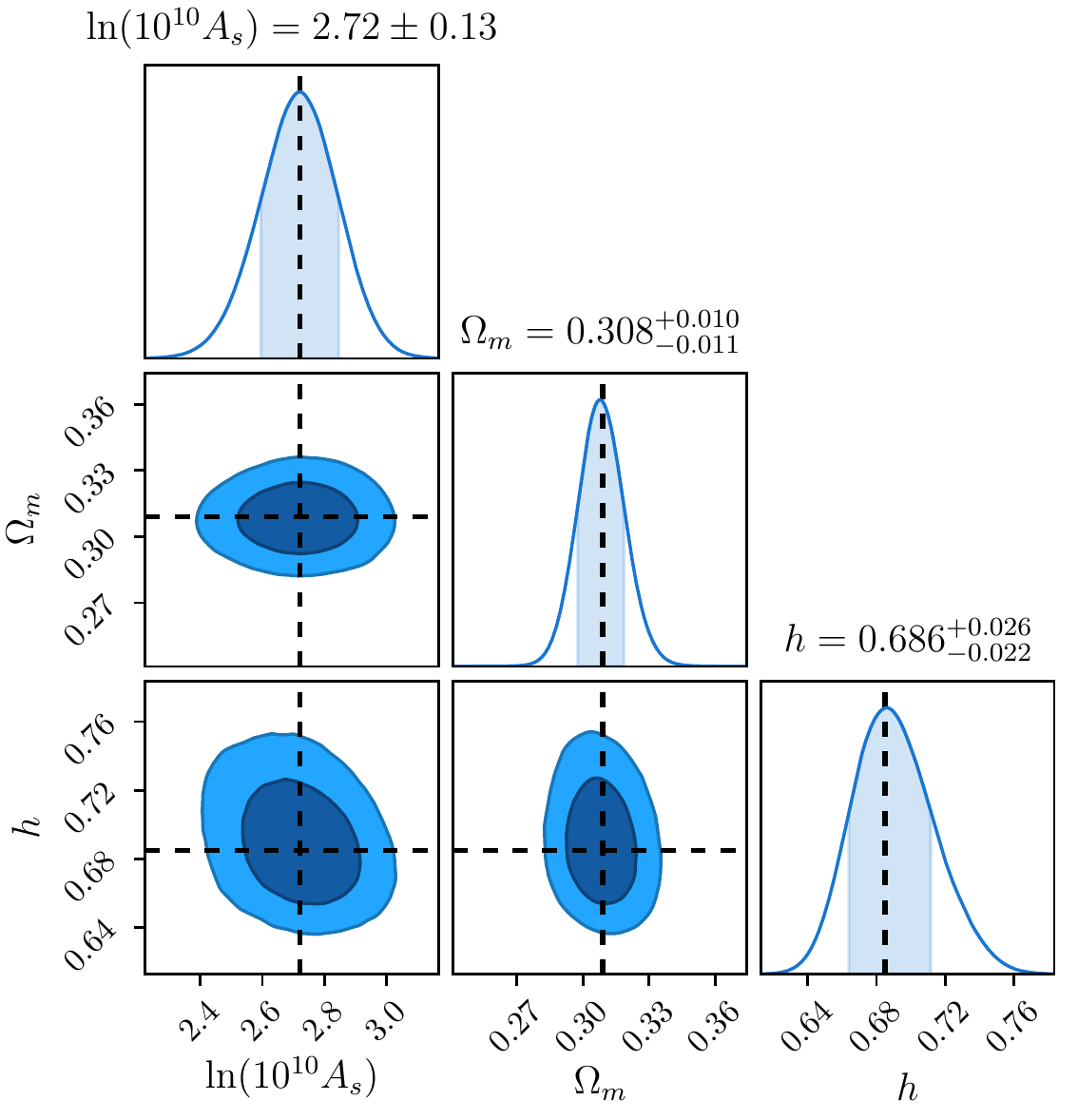}
\caption{\small Posterior distributions for the cosmological parameters obtained from the analysis of the CMASS and LOWZ NGC samples using the Taylor expansion approximation for the EFT power spectrum up to $\kmax=0.20\hinvMpc$ for CMASS and $\kmax=0.18\hinvMpc$ for LOWZ NGC. We put a Planck2018 prior on $f_{bc}$ and on $n_s$ and we fix the neutrino spectrum to {$\sum_i m_{\nu_i} = 0.06$eV with NH.}
In vertical dashed we plot the expectation value from~\cite{DAmico:2019fhj}. } 
\label{fig:analysis_fixed_nu}
\end{figure}

We are now ready to analyze the BOSS DR12 data. We focus on the monopole and quadrupole and analyze the data up to $\kmax=0.2\hinvMpc$ for the CMASS sample and up to $\kmax=0.18\hinvMpc$ for the LOWZ sample.  First, similarly to what Planck2018 did and also to what was done in~\cite{DAmico:2019fhj}, we impose a normal hierarchy with sum of masses equal to $0.06$ eV.
We impose a Planck2018 prior on~$f_{bc}$: a Gaussian prior with $f_{{bc}, \rm center}=0.1860$ and $\sigma_{f_{{bc}, \rm prior}}=0.0031$; and on $n_s$: a Gaussian prior with $n_{s,\rm center}=0.9649$ and $\sigma_{n_{s},\rm prior}=0.0044$. The results of the analysis are shown in Fig.~\ref{fig:analysis_fixed_nu} and Table~\ref{tab:summarydata}.
We find that the central values and error bars for $A_s, \Omega_m$ and $h$ are extremely similar to the ones of~\cite{DAmico:2019fhj} (see Table 4 there, the difference is less than {$\sigma_{\rm stat}/8$}). This shows that having fixed $f_{bc}$ and $n_s$ to the Planck2018 data and $\sum_i m_{\nu_i}$ to 0.06 eV was a good approximation in place of imposing Planck2018 priors on the same parameters: the actual Planck2018 posteriors for these parameters are sufficiently narrow with  respect to what could be measured with our data.  This result has also the consequence that we do not need to test against simulations the presence of a significant  theoretical systematic error  in the case under consideration. In~\cite{DAmico:2019fhj} a careful comparison  with several sets of simulations was performed by fixing $f_{bc}$ and $n_s$ to the Planck2018 preferred values, and it was found that the theoretical systematic error of the EFTofLSS is negligible up to $\kmax=0.2\hinvMpc$ for the BOSS data. Given that the priors that we impose on $f_{bc}$ and $n_s$ are so small that the value of the freely ranging parameters are hardly affected, we conclude that the theoretical systematic error is still sufficiently small to be neglected. For the same reasons, the physical interpretation of the results and physical considerations are the same as~\cite{DAmico:2019fhj}, to which we refer.

\begin{table}[!]
\centering
\scriptsize
\begin{tabular}{ |  c | c | c | c | c | c | c | c |} \hline
Prior & $\lnAs$ & $\Omega_m$ & $h$  & $f_{bc}$ & $n_s$ & $\omega_b$ & $\sum m_\nu$ [eV]               \\  \hline 
fixed $f_{bc}$\,\&\,$n_s$\, \&\, $\nu$    & $2.72\pm 0.12$ & $0.308\pm 0.010$   & $0.686\pm 0.023$ & Prior & Prior & --- & $0.06$  \\
Planck2018, fixed $\nu$   & $2.72\pm 0.13$ & $0.308\pm 0.010$   & $0.686\pm 0.024$ & Prior & Prior & --- & $0.06$   \\
Planck2018, $\sum m_\nu\leq0.9$eV    & $2.77\pm 0.12$ & $0.316\pm 0.012$ & $0.716\pm 0.031$  &  Prior & Prior & --- & $0.24^{+0.30}_{-0.14}$ \\
Planck2018, $\sum m_\nu\leq0.25$eV   & $2.75\pm 0.12$ & $0.310\pm 0.011$ & $0.692\pm 0.025$ &  Prior & Prior & --- & Prior  \\
Planck2018, $\sum m_\nu\leq0.9$eV   & $2.81 \pm 0.13$ & $0.322\pm 0.014$ & $0.690\pm 0.014$ & --- & Prior & Prior & $0.31^{+0.27}_{-0.17}$  \\
BBN, $\sum m_\nu\leq0.9$eV   & $2.91 \pm 0.19$ & $0.314\pm 0.018$ & $0.687\pm 0.015$ & --- &  {$0.979^{+0.075}_{-0.068}$ }& Prior & $0.29^{+0.31}_{-0.18}$  \\
BBN, $\sum m_\nu\leq0.25$eV   & $2.79 \pm 0.16$ & $0.315\pm 0.018$ & $0.686\pm 0.015$  & --- &  {$0.950^{+0.052}_{-0.062}$ }& Prior & Prior \\
\hline
\end{tabular}
\caption{\small 68\% confidence intervals for the cosmological parameters from the  individual analyses over CMASS sample and LOWZ NGC sample of the BOSS data up to $\kmax = 0.20 \hinvMpc$ for CMASS and $\kmax=0.18\hinvMpc$ for LOWZ NGC, apart for $\kmax=0.23\hinvMpc$ for CMASS and $\kmax=0.20\hinvMpc$ for LOWZ NGC in the last two lines. The entry `Prior' allows us to identify on which parameters the prior dominates the constraint.
\label{tab:summarydata}}
\end{table}

\subsection{Free-varying Neutrino spectrum}

\subsubsection{Planck prior on $f_{bc}$ and $n_s$}

\begin{figure}[h!]
\centering
  \begin{center}
  \textbf{Planck prior on  $\Omega_b/\Omega_c$ and $n_s$}
    \end{center}
\includegraphics[width=0.495\textwidth,draft=false]{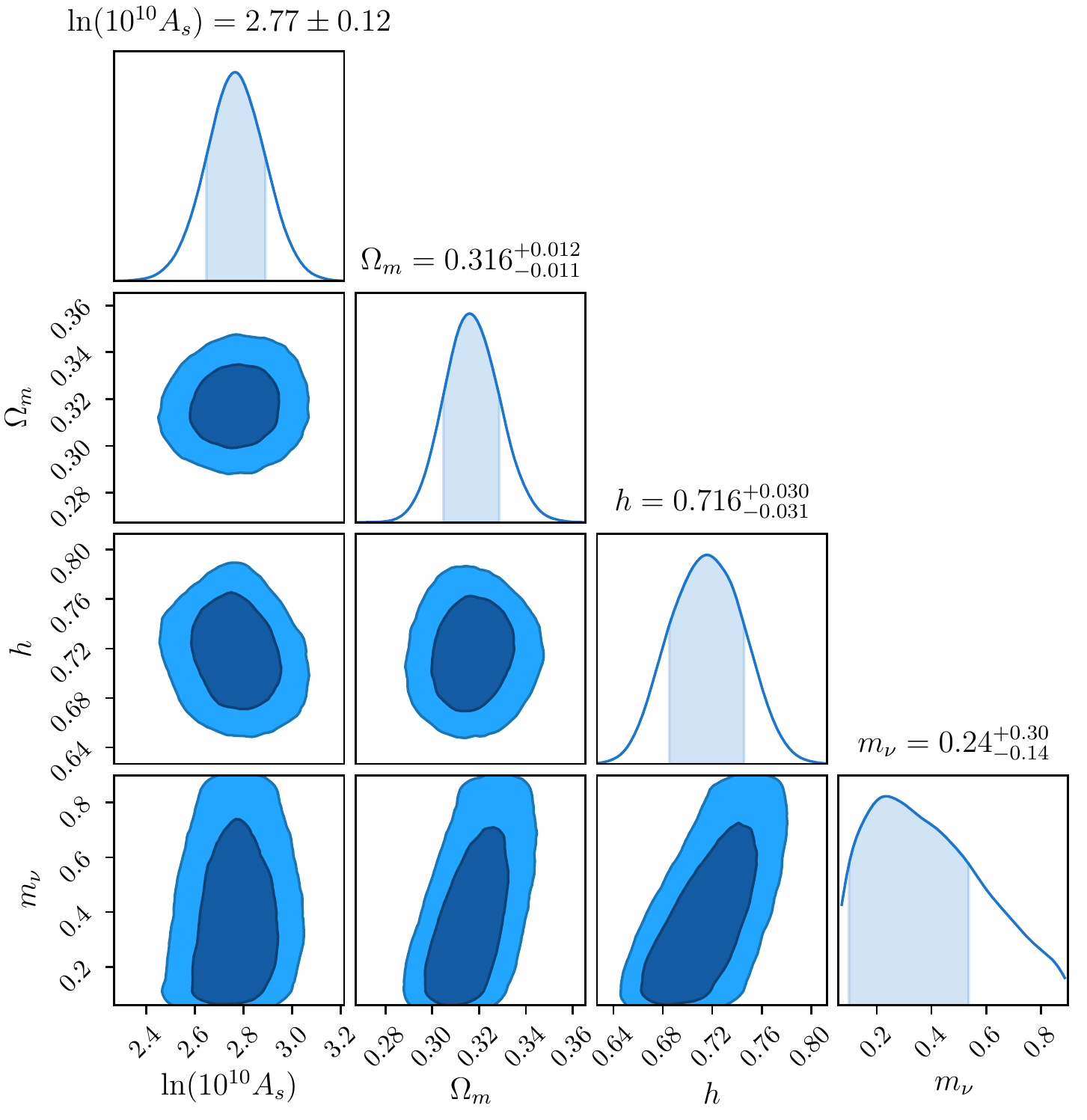}
\includegraphics[width=0.495\textwidth,draft=false]{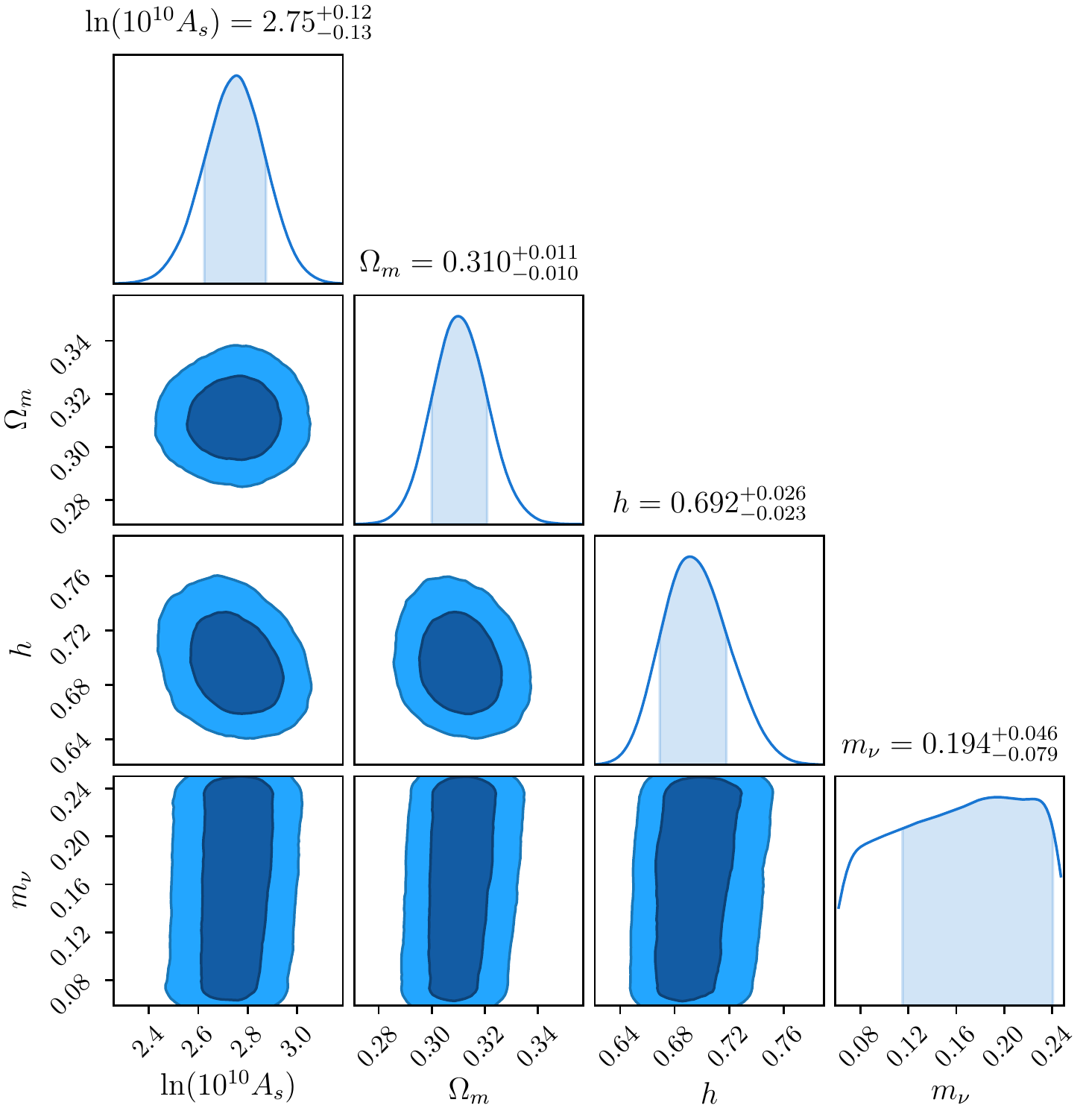}
\caption{\small {\it Left:} Posterior distributions for the cosmological parameters obtained from the analysis of the CMASSxLOWZ NGC sample using the Taylor expansion approximation for the EFT power spectrum up to $\kmax=0.20\hinvMpc$ for CMASS and $\kmax=0.18\hinvMpc$ for LOWZ NGC. We put a Planck2018 prior on $f_{bc}=\Omega_b/\Omega_c$ and $n_s$. {\it Right:} Same as the left plot, but with an additional flat prior for the neutrino masses: $0.06\, {\rm eV}\leq \sum_{i} m_{\nu_i}\leq 0.25\,{\rm eV}$.}
\label{fig:analysis_total}
\end{figure}

We now analyze the same data by letting the sum of the neutrino masses vary within the range $[0.06,0.9]\,$eV, while imposing always a normal hierarchy. We work up to $\kmax=0.20\hinvMpc$. The effects of neutrinos are approximated in exactly the same way, and with the same caveats, as discussed in sec.~5.2 of~\cite{DAmico:2019fhj}, to which we refer the reader for details. This method should correctly account for the leading effects. The results of the analysis are given in Fig.~\ref{fig:analysis_total} and in Table~\ref{tab:summarydata}. We see that as anticipated in~\cite{DAmico:2019fhj}, $A_s$ is moved to higher values, closer to the central value from Planck2018, but also $\Omega_m$ and $h$ are slightly shifted to higher values, as all of these parameters are positively correlated with neutrino masses. Overall, neutrinos are bounded to be lighter than {$0.78$ eV} at 95\% C.L., which is consistent with the upper bound from Planck2018~\cite{Aghanim:2018eyx}.

On the right of Fig.~\ref{fig:analysis_total}, we plot the same analysis but by adding a flat prior on the sum of neutrino masses: $0.06\, {\rm eV}\leq \sum_{i} m_{\nu_i}\leq 0.25\,{\rm eV}$, motivated  by Planck2018~\cite{Aghanim:2018eyx}. One notices the slight shift in the cosmological parameters implied by this analysis.

\subsubsection{Planck prior on $\omega_b$ and $n_s$}
\begin{figure}[h!]
\centering
  \begin{center}
  \textbf{Planck prior on  $\omega_b$ and $n_s$}
    \end{center}
\includegraphics[scale=0.75,draft=false]{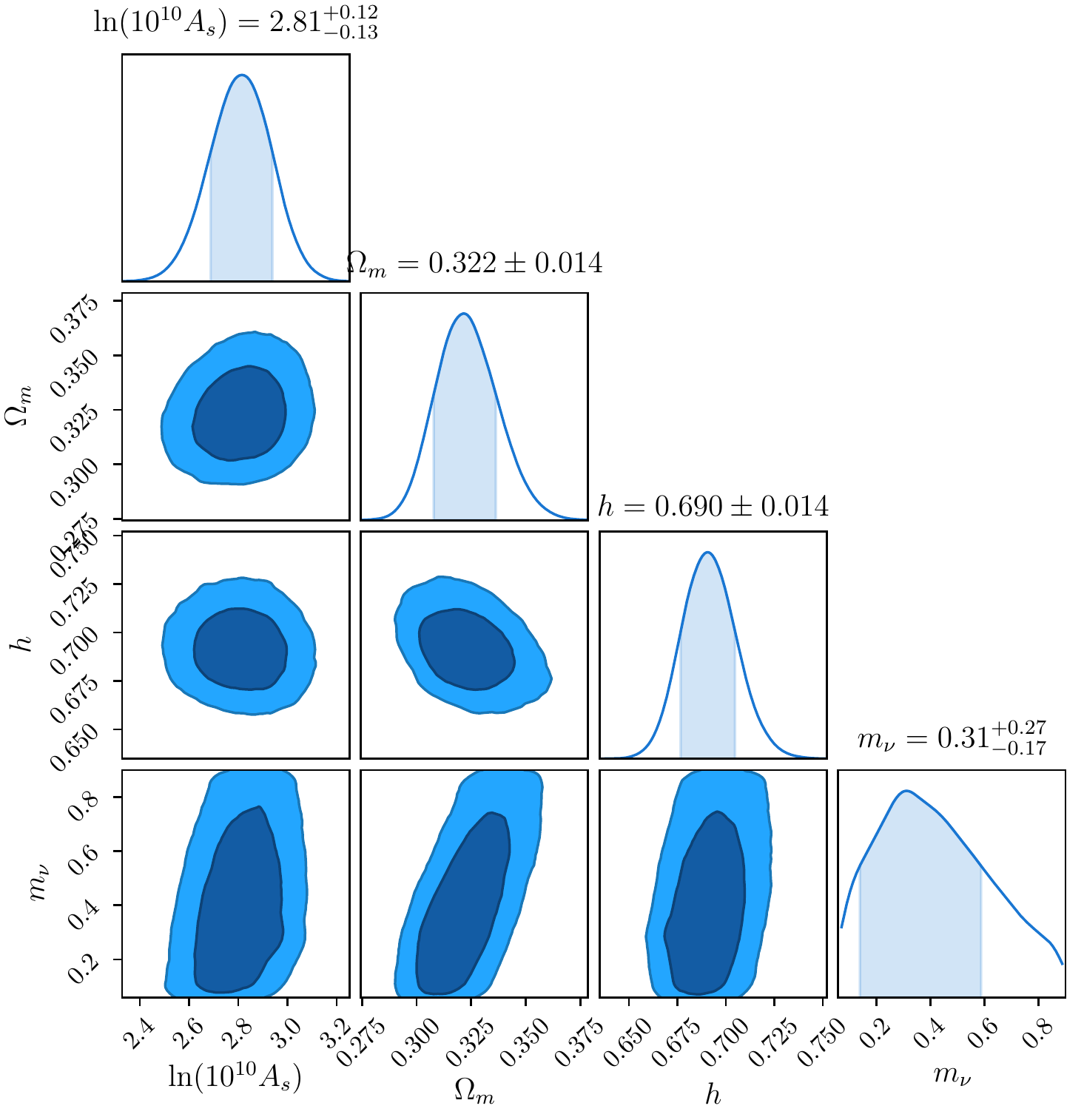}
\caption{\small Posterior distributions for the cosmological parameters obtained from the analysis of the CMASS sample combined with the LOWZ NGC sample, using the Taylor expansion approximation for the EFT power spectrum up to $\kmax=0.20\hinvMpc$ for CMASS and {$\kmax=0.18\hinvMpc$} for LOWZ NGC. We put a Planck2018 prior on $\omega_b=\Omega_b h^2$ and $n_s$.}
\label{fig:analysis_total_new_prior}
\end{figure}

So far, we have always put a prior on $f_{bc} = \omega_b / \omega_c$.
If instead we impose a Planck2018 prior on $\Omega_b h^2$, while keeping a Planck2018 prior on $n_s$ and $0.06\, {\rm eV}\leq \sum_{i} m_{\nu_i}\leq 0.9\,{\rm eV}$, we find the constraints on $A_s,\, \Omega_m$ and $h$ given in Fig.~\ref{fig:analysis_total_new_prior}. One sees that while the error bar on $\Omega_m$ is marginally increased and the one on $A_s$ is also approximately unaffected, the one for $h$ is reduced by about a factor of two. Since $\Omega_b h^2$ is very well measured both by Planck2018 and also from BBN, this suggests that imposing such a prior could be a very interesting way to analyse the BOSS data. Given that the error bars on $h$ are so significantly reduced, in this case we cannot trust the analysis of the systematic errors of~\cite{DAmico:2019fhj}, because the minimally detectable error in that analysis is too large in this case.
We therefore do the analysis of the Challenge simulations and of the Patchy mocks, as done in~\cite{DAmico:2019fhj}, to verify that there is no large systematic theory error.
Our results are shown in appendix~\ref{sec:appendix}, where we verify that the theory systematic error is under control even with smaller statistical errors on $h$, by using the data up to $\kmax=0.20\hinvMpc$ for the CMASS sample (and, following the procedure described in~\cite{DAmico:2019fhj}, up to $\kmax=0.18\hinvMpc$ for LOWZ), which are the ones we use in the data analysis.

\subsubsection{BBN prior on $\omega_b$}
\begin{figure}[h!]
\centering
  \begin{center}
  \textbf{BBN prior}
    \end{center}
\includegraphics[scale=0.65,draft=false]{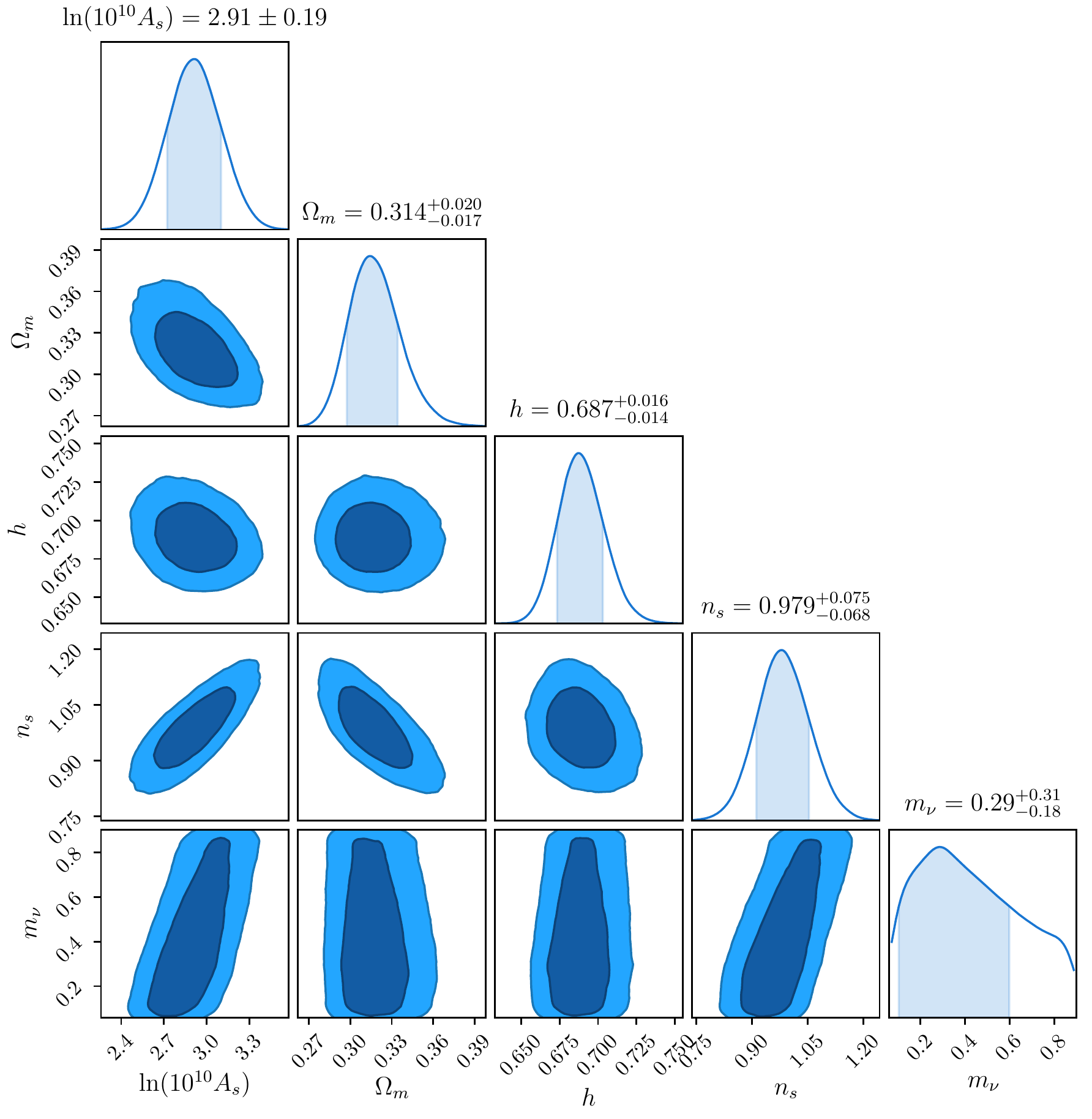}
\caption{\small Posterior distributions for the cosmological parameters obtained from the analysis of the CMASS sample combined with the LOWZ NGC sample, using the Taylor expansion approximation for the EFT power spectrum. We put a BBN prior on $\omega_b=\Omega_b h^2$, working at $\kmax=0.23\hinvMpc$ for CMASS and {$\kmax=0.20\hinvMpc$} for LOWZ NGC. }
\label{fig:analysis_loose_prior}
\end{figure}

If we want to constrain the cosmological parameters using only galaxy clustering and no CMB information, apart for a mild prior on the sum of the neutrino masses, which plays a marginal role, we can change our prior on $\omega_b$.
Namely, we choose a Gaussian prior on $\omega_b$ with $\sigma = 0.00745$ and central value 0.00236, which is dictated by BBN constraints~\cite{Tanabashi:2018oca}~\footnote{Notice that we take the central value to be the one of Planck2018, but the error bar is the one from BBN. The difference is less than $0.2\sigma$.}, for neutrino masses a flat prior $0.06\, {\rm eV}\leq \sum_{i} m_{\nu_i}\leq 0.9\,{\rm eV}$ and no other prior on the remaining cosmological parameters.
In Fig.~\ref{fig:analysis_loose_prior} we use this prior to analyze the combination of CMASS and LOWZ NGC samples. We see that, even without the stringent CMB priors, the cosmological parameters are recovered to a remarkable precision. In particular we notice that the tilt $n_s$ is measured to about 7\% accuracy. 

\begin{figure}[h!]
\centering
\includegraphics[width=0.695\textwidth,draft=false]{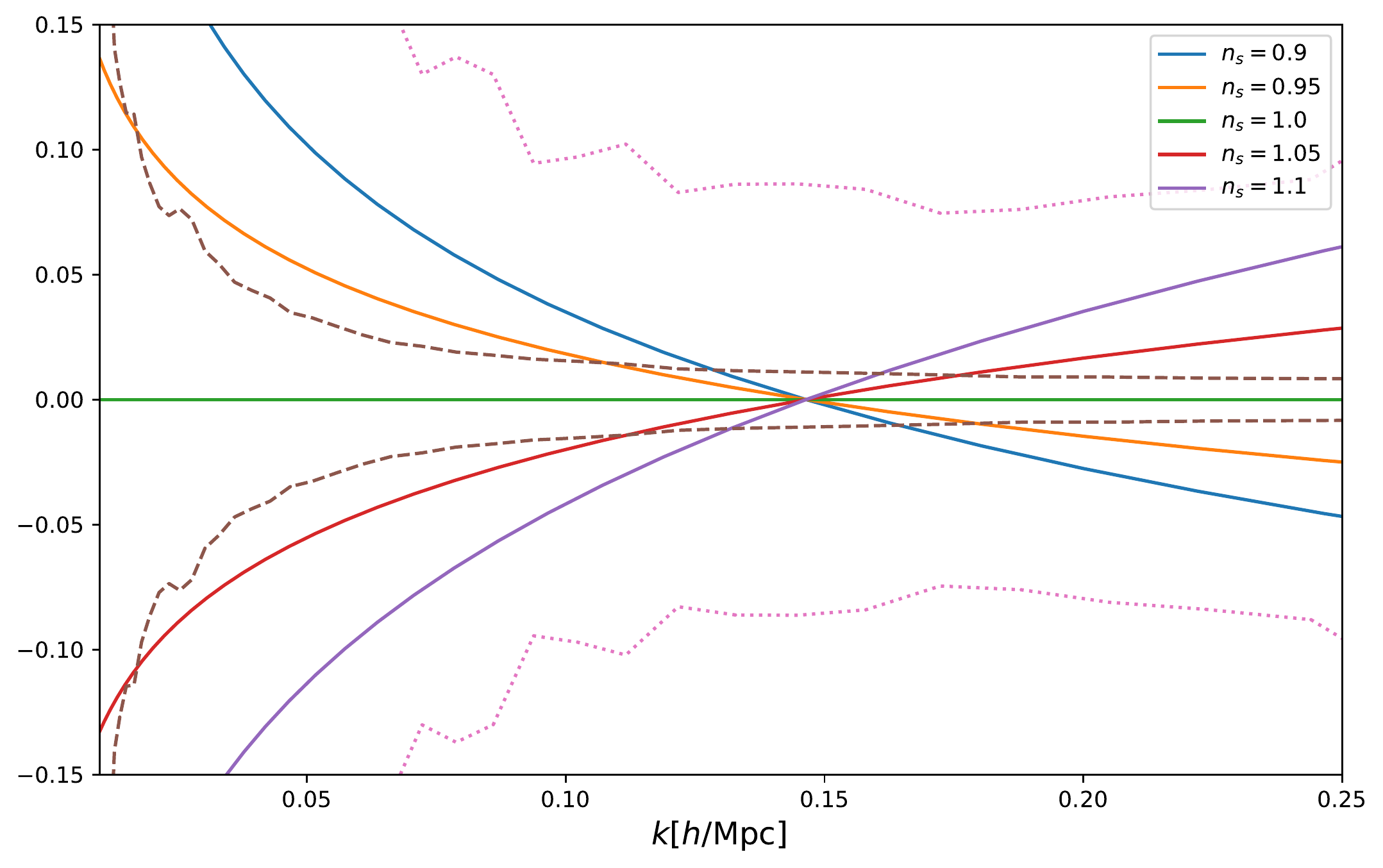}
\caption{\small  Ratio of several linear power spectra  over a power spectrum with $n_s=1.0$, scanning over the tilt, and normalized to agree at $\kmax=0.15\hinvMpc$. Such a normalization accounts for the fact that $A_s$ and the linear bias, $b_1$, can be chosen to match the high-$k$ amplitude of the power spectrum. By $\kmax=0.15\hinvMpc$, loop terms have become smaller, and the EFT-parameters have been determined to fit the high-$k$ wavenumbers, so that degeneracies between loops and tilt should play no major role. We also plot the error bars for the monopole in brown dashed, and the quadrupole, in pink dotted. We can see that deviations from scale invariance of order $5\%$, or even less if there were no degeneracies, should be detectable, as we find in our analysis.}
\label{fig:tilt_tetectable}
\end{figure}

The fact that the data are powerful enough to measure the tilt to this accuracy can be understood in the following way. In Fig.~\ref{fig:tilt_tetectable} we plot ratios of several linear power spectra over a power spectrum with $n_s=1.0$, normalized in such a way to agree at $\kmax=0.15\hinvMpc$. The normalization at $\kmax=0.15\hinvMpc$ is chosen so that one, implicitly, assumes that $A_s$ and the linear bias, $b_1$, have been chosen to fit the data at high wavenumbers. We did not choose to normalize the power spectra at higher wavenumbers, such as $\kmax=0.23\hinvMpc$, to account for the fact that the highest wavenumbers might be used to determine all the EFT-parameters, in such a way that degeneracies with the loop contributions (which at $\kmax=0.15\hinvMpc$ have become smaller anyway) should play a smaller role. Overall, we see that deviations of the tilt from scale invariance of order $5\%$, or even smaller if there were not  residual degeneracies, should be detectable, in rough agreement with our findings.

In App.~\ref{app:CMASS} we show the results of the analysis of the CMASS sample, showing that the data sets are compatible. In the same appendix,  we show the result of the same analysis but by imposing an additional flat prior for the neutrino masses: $0.06\, {\rm eV}\leq \sum_{i} m_{\nu_i}\leq 0.25\,{\rm eV}$, motivated  by Planck2018~\cite{Aghanim:2018eyx}.  As we show in App.~\ref{sec:appendix}, because of the slightly larger statistical errors,  the theoretical systematic error is negligible even at $\kmax=0.23\hinvMpc$ for CMASS (and, following the procedure described in~\cite{DAmico:2019fhj}, up to $\kmax=0.20\hinvMpc$ for LOWZ), which  is the $\kmax$ at which we perform the analysis in this case.

\section{Conclusions}

After Taylor expanding the predictions of the Effective Field Theory of Large-Scale Structure (EFTofLSS) with respect to the cosmological parameters, and after establishing the accuracy of such an approximation, we have measured the cosmological parameters from the CMASS and LOWZ NGC samples of the BOSS DR12 data. We have done this by matching the cosmology-dependent predictions for the monopole and quadrupole power spectra of galaxies in redshift space from the EFTofLSS against the measured data up to $\kmax=0.2\hinvMpc$ and $\kmax=0.23\hinvMpc$. We have allowed the amplitude of the primordial power spectrum, $A_s$, the abundance of dark matter, $\Omega_m$, and the present value of the Hubble rate, $H_0$, to vary freely, while we have imposed several sets of priors on the other cosmological parameters: either  Planck2018 priors on the ratio of baryons with respect to dark matter, $f_{bc}$, and on the tilt of the primordial power spectrum, $n_s$; or Planck priors on the the baryon fraction of the energy density, $\Omega_b h^2$, and on $n_s$; or a BBN prior on $\Omega_b h^2$ {while leaving $n_s$ free}. For the neutrinos, by assuming a normal hierarchy, we have either allowed their total mass to vary within a large range of $[0.06, 0.9]$eV, or fixed it to $0.06$eV, or even imposed a Planck-motivated flat prior for this quantity to be smaller than $0.25$eV. Though we have stressed that the modeling of neutrinos is not very accurate, our implementation is expected to account for the leading effects. We have either used the results of~\cite{DAmico:2019fhj} or directly compared with several sets of simulations to conclude that the theoretical systematic error induced by the EFTofLSS modeling is negligibly small.

In summary, we find that, using the CMASS and LOWZ NGC samples of the BOSS DR12 data {up to $\kmax=0.23\hinvMpc$ for CMASS and up to $\kmax=0.20\hinvMpc$ for LOWZ NGC}, and by imposing only a BBN prior on $\Omega_b h^2$ and a flat prior $0.06\, {\rm eV}\leq \sum_{i} m_{\nu_i}\leq 0.9\,{\rm eV}$,  we can measure $A_s$ to 19\%, $\Omega_m$ to 5.7\%, $H_0$ to 2.2\%, $n_s$ to 7.3\% accuracy, and bound the neutrinos to be lighter than 0.83 eV at 95\% confidence level. The $68\%$ confidence level intervals for $A_s,\,\Omega_m$, $H_0$ and $n_s$ read $\lnAs=2.91\pm 0.19$, $\Omega_m=0.314\pm 0.018$, $H_0=68.7\pm 1.5$ km/(s Mpc) and $n_s=0.979\pm 0.071$. Similar results hold for different sets of priors. 

With respect to the analysis performed in~\cite{DAmico:2019fhj}, the main generalization of our analysis is the fact that we do not fix $f_{bc}$ and $n_s$ to the best values from Planck2018, but rather impose a consistent prior, and also that we consider the case of a prior on $\Omega_b h^2$ instead of $f_{bc}$, including a rather wide one that relies only marginally on the CMB and does not constrain $n_s$. In all these cases, our Monte Carlo Markov Chain (MCMC) needs to explore a much-higher dimensional  cosmological parameter space. In order to do this in an effective way, following the idea initially developed in~\cite{Cataneo:2016suz}, we have used the fact that the range of parameter space that is consistent with the observational data is small enough to approximate the dependence of the EFTofLSS power spectrum on cosmological parameters with a Taylor expansion around a reference cosmology. {By comparison with direct calculations of the same observable, we have found that, for the data we analyze, it is enough to include all the third order derivatives (though this is probably a conservative requirement, see footnote~\ref{footnote:lower_conv}).} The resulting power spectrum and likelihoods can be evaluated extremely fast. Additionally, the Taylor-expanded prediction can be constructed with the evaluation of a small number of power spectra, making the overall procedure very expedite. We publicly release the whole pipeline  of the analysis with the appearance of this preprint.\\

There are several directions that our findings open up:
\begin{itemize}

\item It would be interesting to remove the BBN prior on $\Omega_b h^2$, and see what are the constraints that the BOSS observations can impose {\it per se}.

\item It should be possible to include additional cosmological parameters and perform an analysis exploring an even larger parameter space. For example, one could include non-vanishing curvature, primordial non-Gaussianities, an equation of state for dark energy, dynamical  dark energy, etc., as well as a more accurate implementation of the effect of neutrino masses, as described in~\cite{DAmico:2019fhj}. 
\end{itemize}

We leave this, and more, to future work. More generally, we believe this work, together with~\cite{DAmico:2019fhj}, shows the usefulness and versatility of using the EFTofLSS to analyze Large-Scale  Structure data, which, in turn, are revealed to be extremely rich in cosmological information.

\section*{Acknowledgements}

We thanks H\'ector Gil-Mar\'in for support with the BOSS data.
TC and PZ would like to express their gratitude to the people of SLAC and SITP for their warm welcome where part of this work was done.
LS and GDA are partially supported by Simons Foundation Origins of the Universe program (Modern Inflationary Cosmology collaboration) and LS by NSF award 1720397. 
The Boltzmann code CLASS was used to compute the matter linear power spectrum~\cite{Blas:2011rf}. The MCMC samples were performed using {\it emcee}~\cite{ForemanMackey:2012ig} Some plots were done using the Chainconsumer package~\cite{Hinton2016}. Part of the analysis was performed on the Sherlock cluster at the Stanford University, for which we thank the support team, and part on the Mercury cluster of the University of Science and Technology of China, for which PZ thanks Xin Ren for support.

\appendix

\section{Further checks for the Taylor Expansion\label{app:furtherchecks}}

\begin{figure}[h!]
\centering
\includegraphics[width=0.47\textwidth,draft=false]{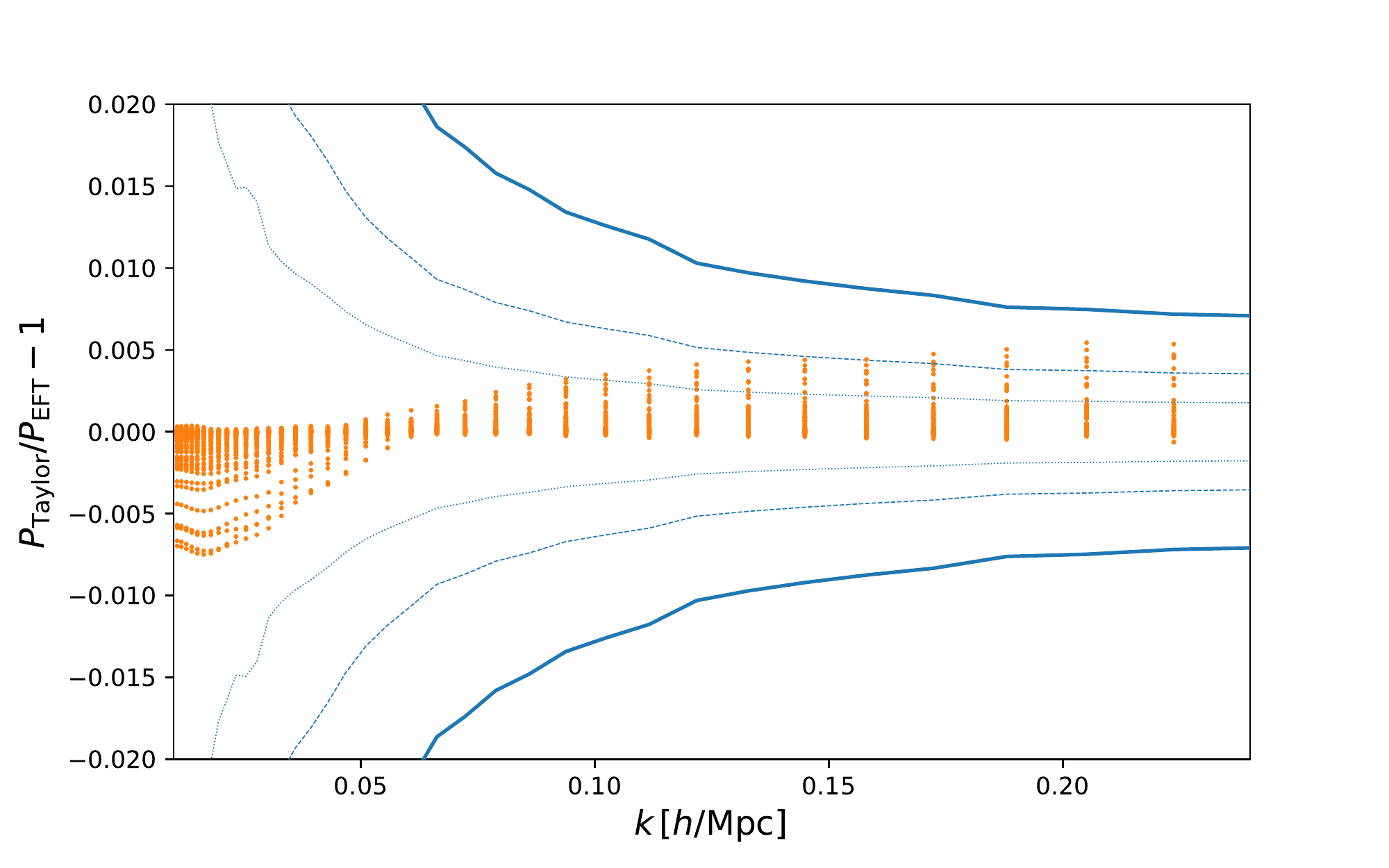}
\includegraphics[width=0.47\textwidth,draft=false]{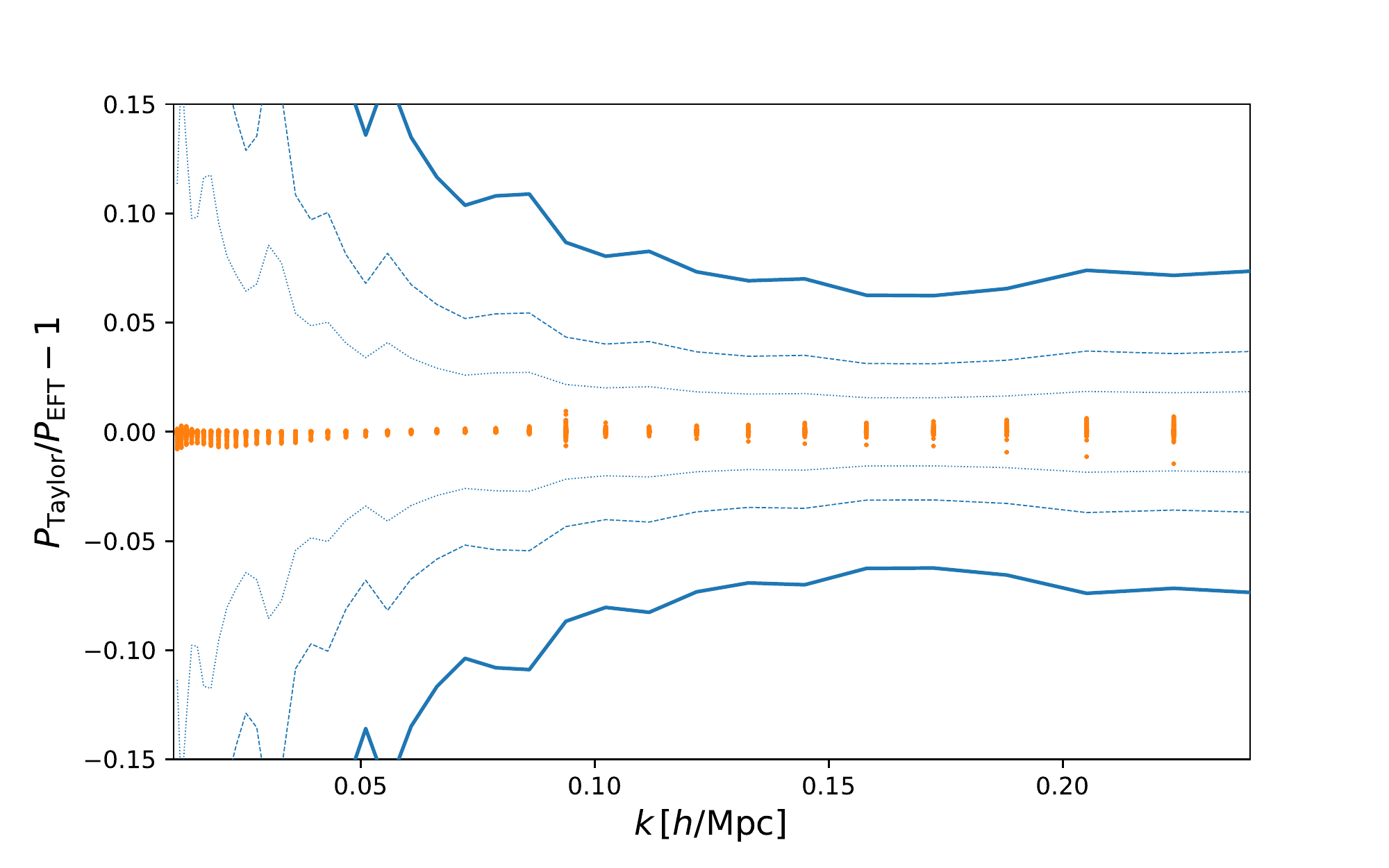}
\caption{\small Relative difference  between the computation of an EFTofLSS power spectrum by the direct evaluation or by approximation with the Taylor expansion, 
for cosmologies that lie on a hyper-ellipsoid with orthogonal semi-axes of length  $3 \sigma_{\rm stat}$, centered on the reference cosmology, where $3 \sigma_{\rm stat}$ are obtained from the analysis of the CMASSxLOWZ sample with the BBN prior on $\omega_b$. 
On the left we plot the monopole, on the right the quadrupole. 
In solid blue, we plot the $1\sigma$ error bars of the CMASS data, and in dashed and dotted blue the $\sigma/2$ and $\sigma/4$ error bars of the data, respectively.
We see that the disagreement is very small when confronted to the error bars of the data even for these statistically disfavored cosmologies. } \label{fig:threesigma}
\end{figure}

\begin{figure}[h!]
\centering
\includegraphics[width=0.495\textwidth,draft=false]{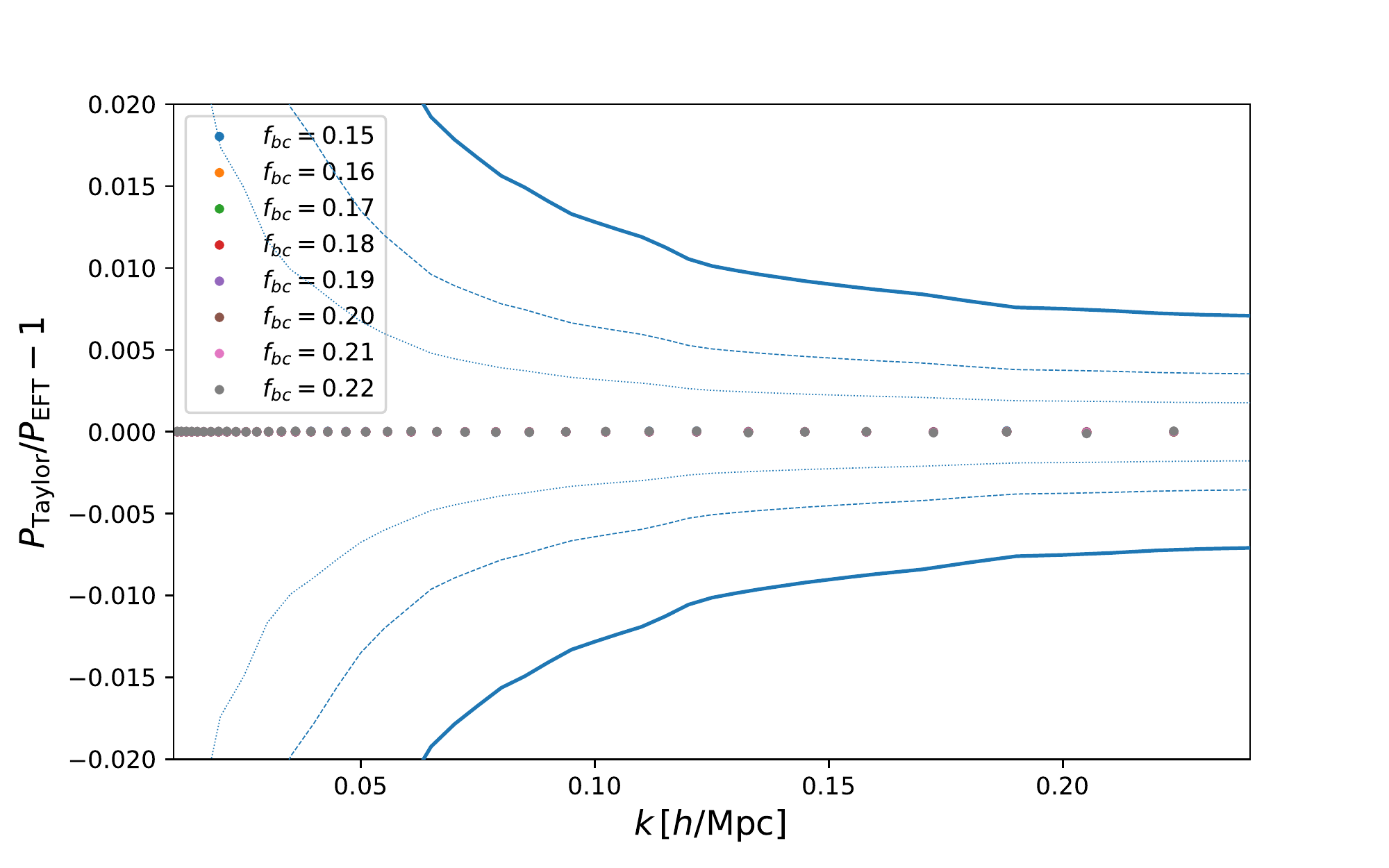}
\includegraphics[width=0.495\textwidth,draft=false]{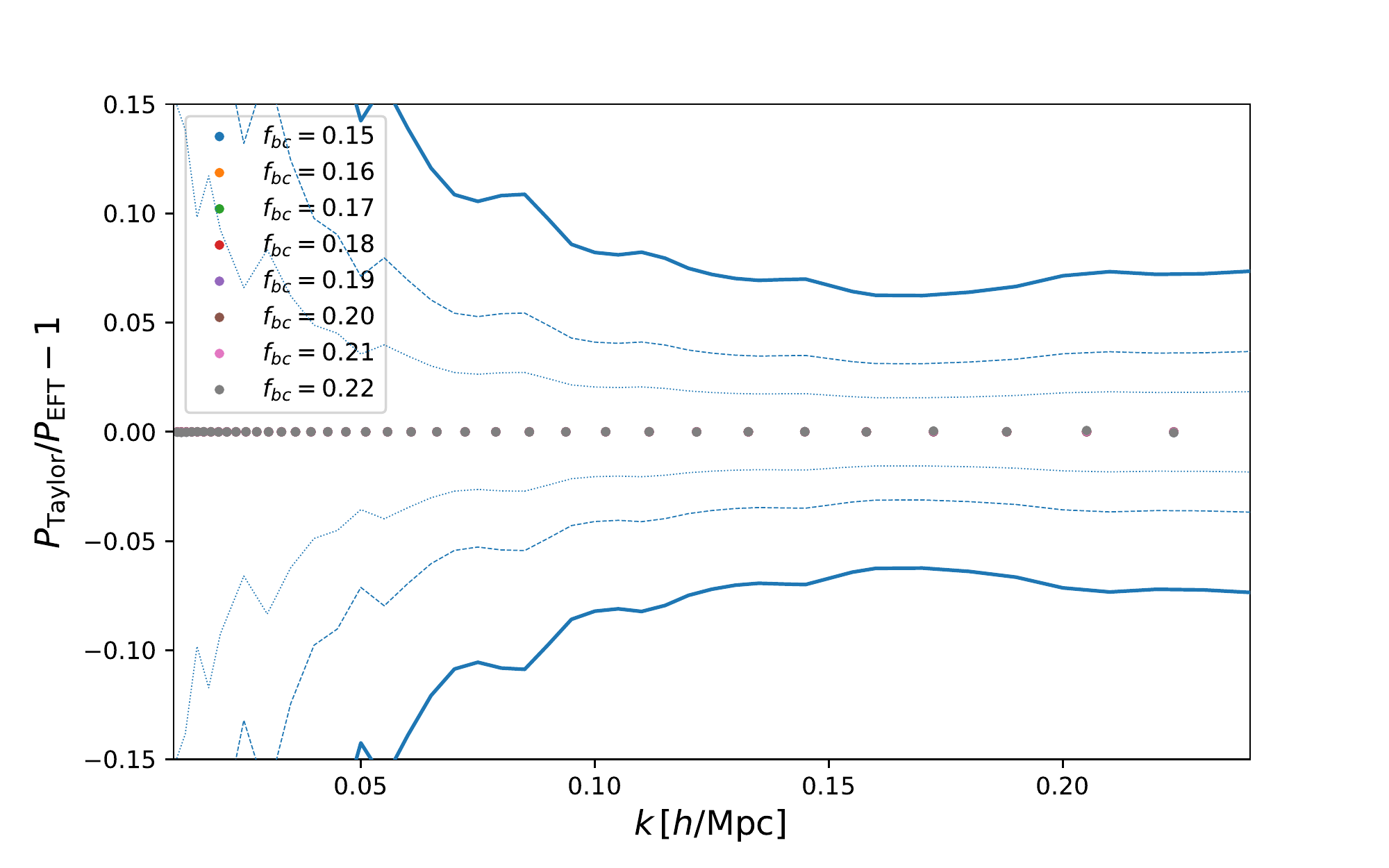}
\caption{\small  Relative difference between the computation of a EFTofLSS power spectrum by the direct evaluation or by approximation with the Taylor expansion, for cosmologies that lie on the fiducial cosmology, in which we vary the ratio $f_{bc}$ by about 40\% from $0.15$ to $0.22$, as indicated in the legend. In solid blue, we plot the $1\sigma$ error bars of the CMASS data, and in dashed and dotted blue the $\sigma/2$ and $\sigma/4$ error bars of the data, respectively. On the left we plot the monopole, on the right the quadrupole. We see that the disagreement is negligibly small when compared to the error bars of the data.}
\label{fig:relativepowerfbc}
\end{figure}

\begin{figure}[h!]
\centering
\includegraphics[width=0.495\textwidth,draft=false]{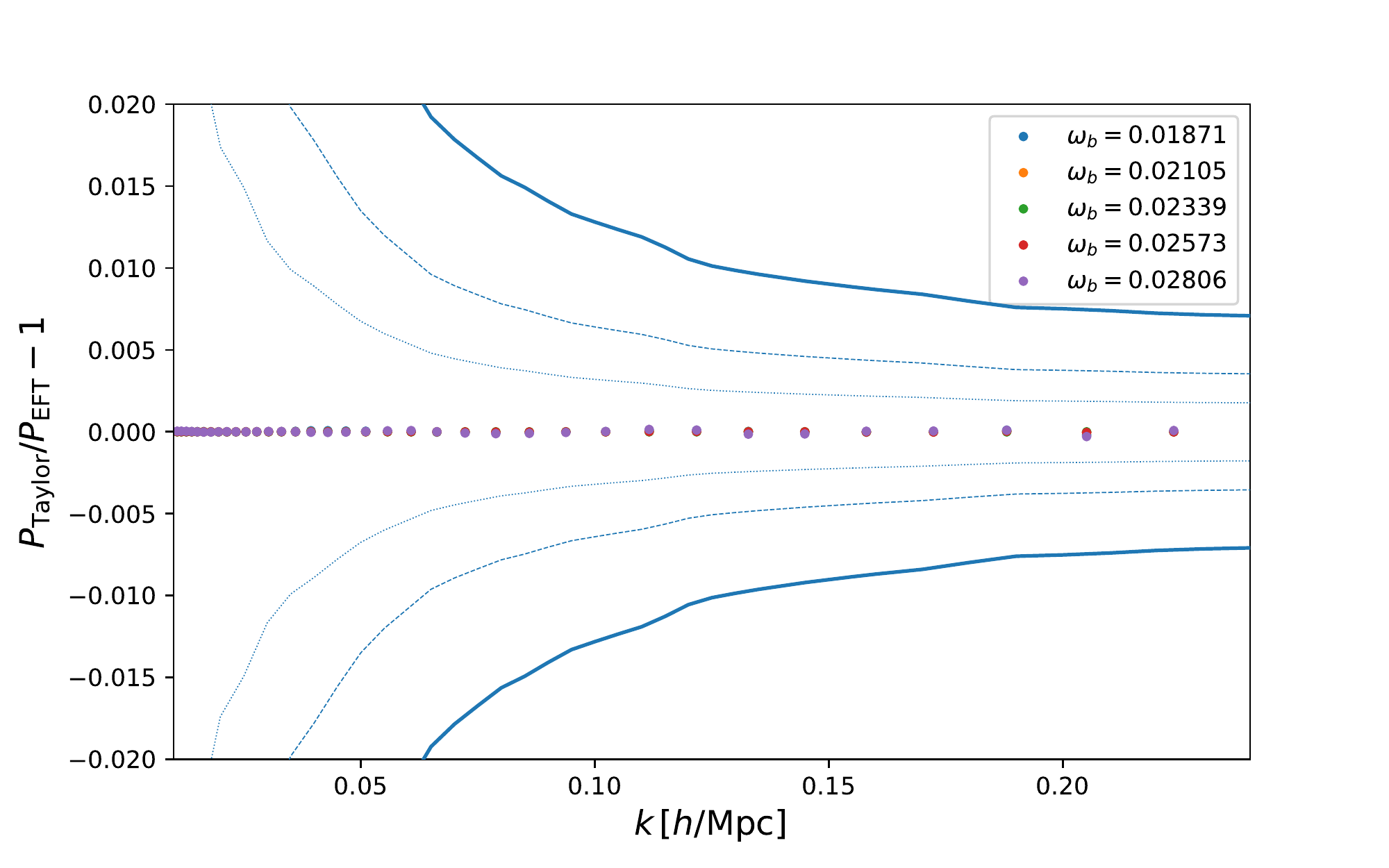}
\includegraphics[width=0.495\textwidth,draft=false]{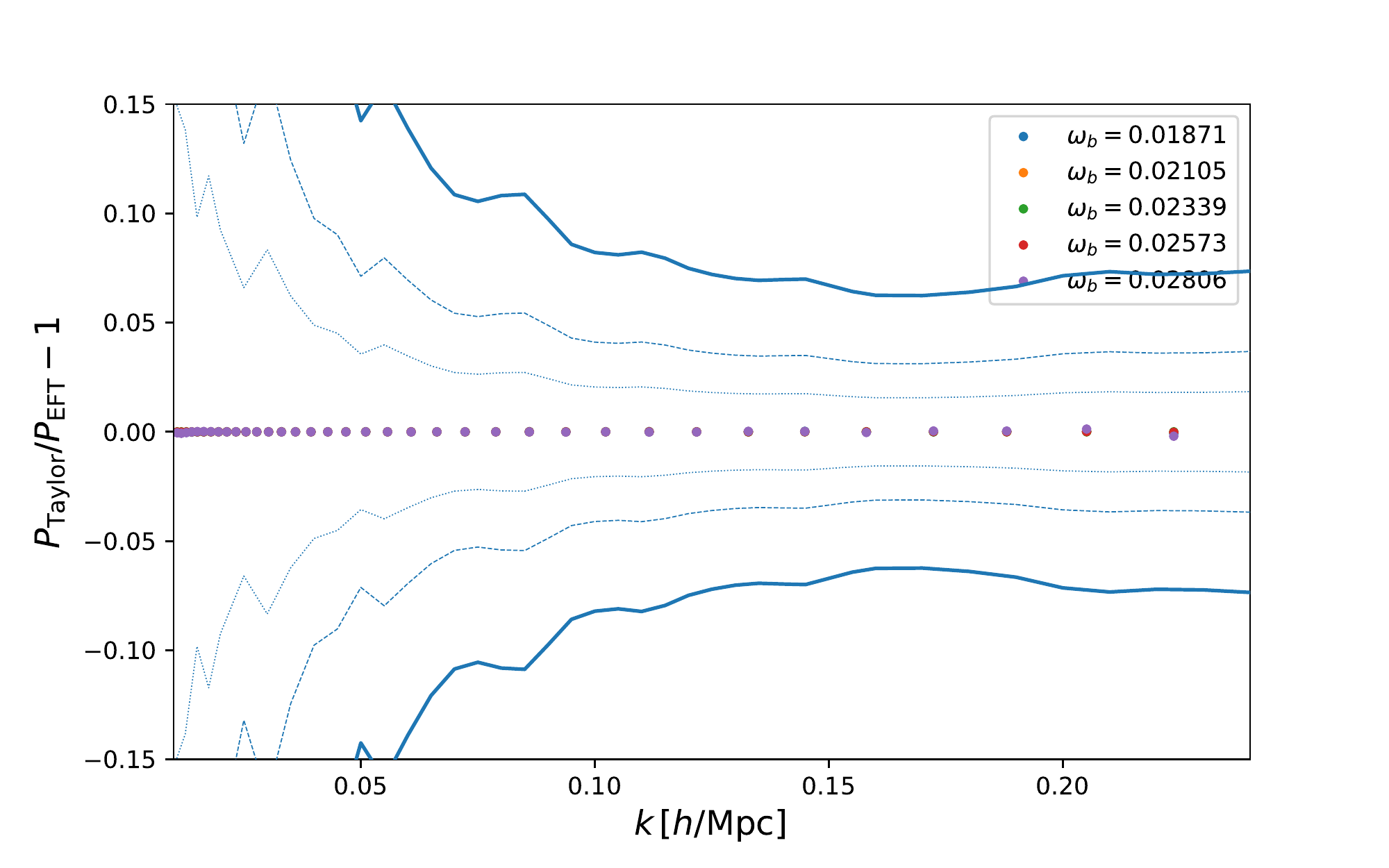}
\caption{\small Relative difference between the computation of a EFTofLSS power spectrum by the direct evaluation or by approximation with the Taylor expansion, for cosmologies that lie on the fiducial cosmology, in which we vary the ratio $\omega_b$ by {40\% from $0.01871$ to $0.02806$}, as indicated in the legend. {In solid blue, we plot the $1\sigma$ error bars of the CMASS data, and in dashed and dotted blue the $\sigma/2$ and $\sigma/4$ error bars of the data, respectively.} On the left we plot the monopole, on the right the quadrupole. We see that the disagreement is negligibly small when compared to the error bars of the data.}
\label{fig:relativepoweromb}
\end{figure}

\begin{figure}[h!]
\centering
\includegraphics[width=0.495\textwidth,draft=false]{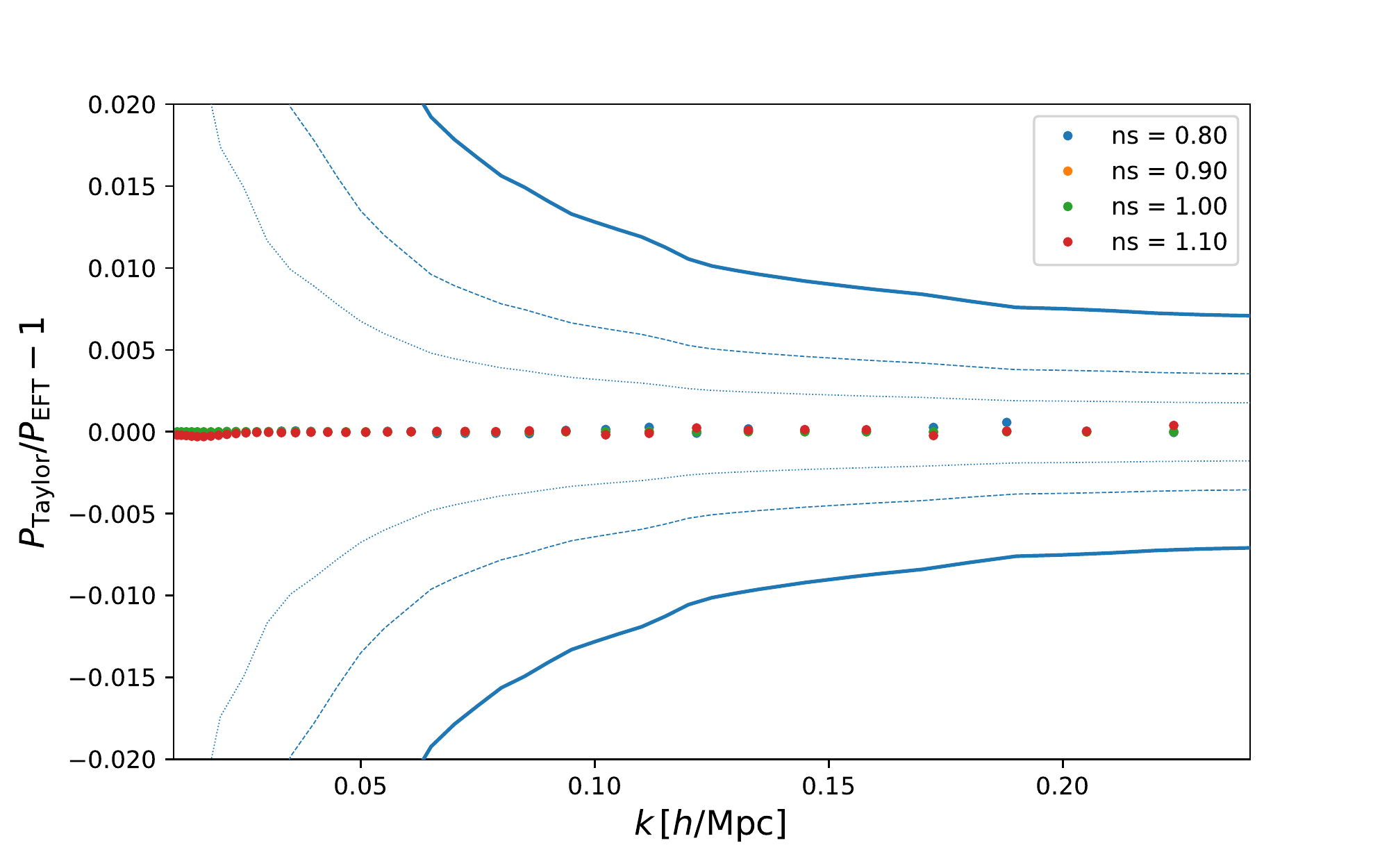}
\includegraphics[width=0.495\textwidth,draft=false]{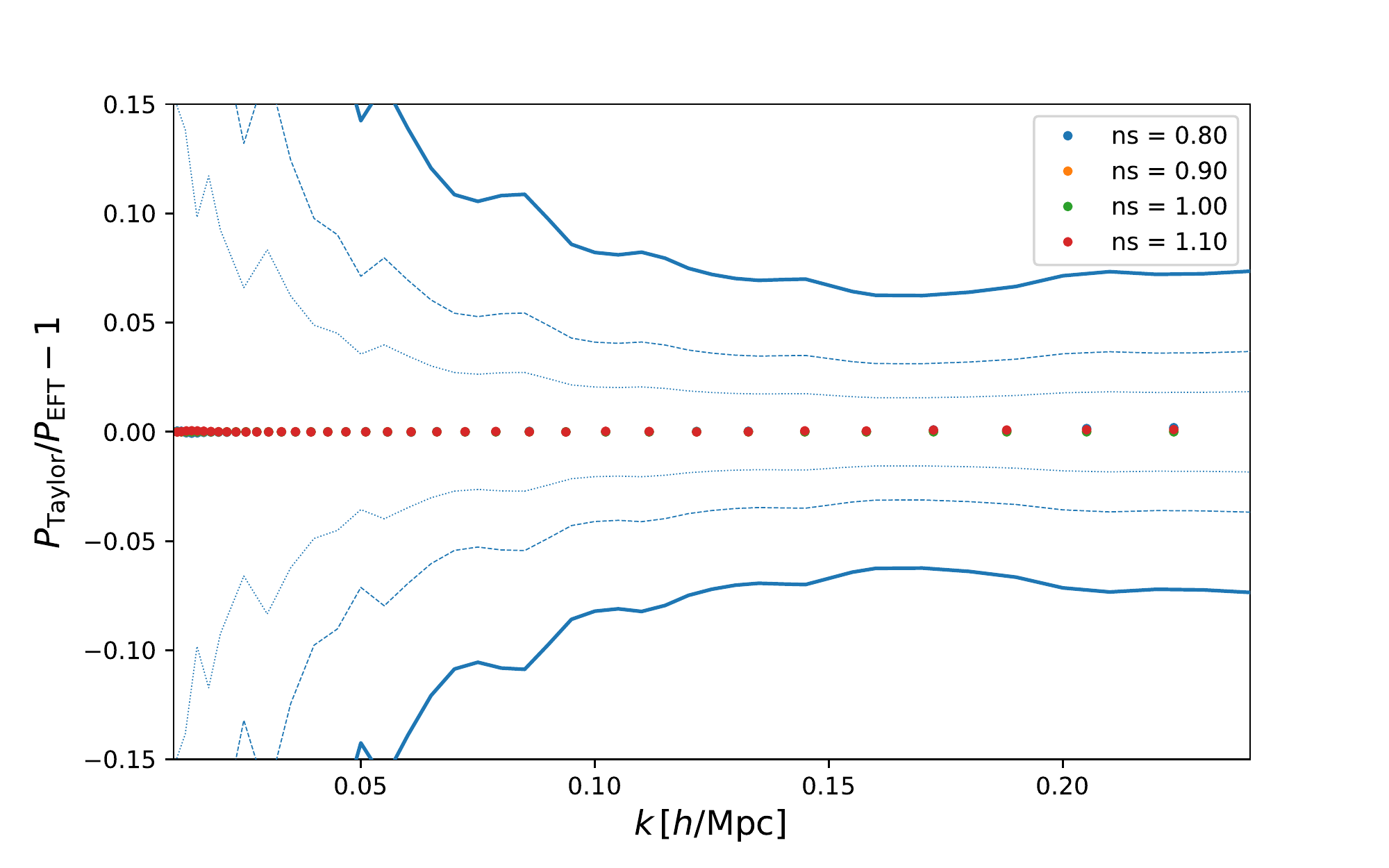}
\caption{\small Relative difference between the computation of a EFTofLSS power spectrum by the direct evaluation or by approximation with the Taylor expansion, for cosmologies that lie on the fiducial cosmology, in which we vary the tilt of the power spectrum, $n_s$, by about 30\% from $0.8$ to $1.1$, as indicated in the legend. {In solid blue, we plot the $1\sigma$ error bars of the CMASS data, and in dashed and dotted blue the $\sigma/2$ and $\sigma/4$ error bars of the data, respectively.} On the left we plot the monopole, on the right the quadrupole. We see that the disagreement is negligibly small when compared to the error bars of the data.}
\label{fig:relativepowerns}
\end{figure}

In this appendix, we show some further checks that we perform to clarify the accuracy of the Taylor expanded approximation for the EFTofLSS prediction.
{In Fig.~\ref{fig:threesigma}, we show the relative difference between the computation of a EFTofLSS power spectrum by the direct evaluation or by approximation with the Taylor expansion, for cosmologies that lie on an hyper-ellipsoid with orthogonal semi-axes of length  $3 \sigma_{\rm stat}$, centered on the reference cosmology, where $3 \sigma_{\rm stat}$ are obtained from the analysis of the CMASSxLOWZ sample with the BBN prior on $\omega_b$ (in particular, $\sum m_\nu$ spans over $\left[0.07, 0.9\right]$eV). }
In Fig.~\ref{fig:relativepowerfbc}, ~\ref{fig:relativepoweromb} and~\ref{fig:relativepowerns}, we show the relative deviation on the reference cosmology varying $f_{bc}$, $\omega_b$ and $n_s$ over a range much wider than theirs Planck2018 or BBN priors or $\sigma_{\rm CMASSxLOWZ}$, respectively by 40\%, 40\% and 30\%. On the same figures, we also plot the $1\sigma$ error bars of the data.
We see that the deviations between the Taylor expansion and  the actual calculation is safely smaller than the error bars and so negligible. 
{These results not only confirm the accuracy of the Taylor expansion for this analysis, but tell us that one can use such numerical approximation to explore much farther cosmologies than the ones probed here.}

\section{Tests on Challenge simulations and Patchy mocks}
\label{sec:appendix}
\begin{figure}[h!]
\centering
    \begin{center}
  \textbf{Planck Priors ABFG \& D}
    \end{center}
\includegraphics[width=0.495\textwidth,draft=false]{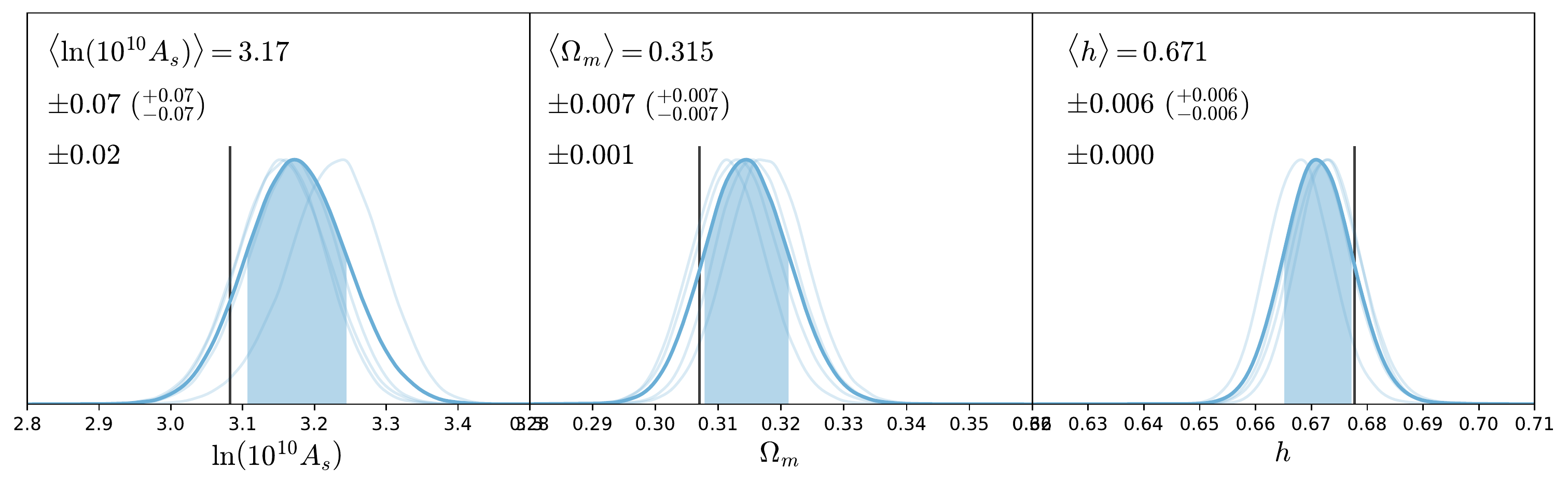}
\includegraphics[width=0.495\textwidth,draft=false]{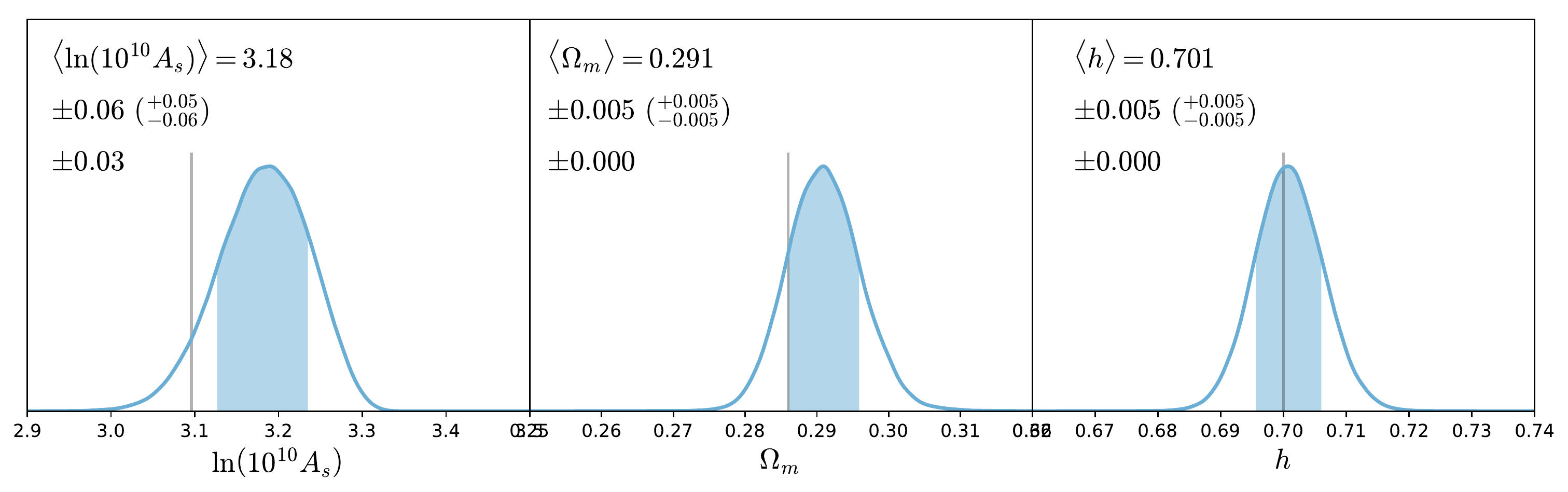}
\caption{\small  Posterior distributions for the cosmological parameters obtained from the analysis of the Challenge simulations A, B, F, G ({\it left}), and D ({\it right}) using the Taylor expansion approximation for the EFT power spectrum, and by imposing the Planck2018 prior on $\omega_b=\Omega_b h^2$ and $n_s$  at $\kmax=0.20\hinvMpc$.}
\label{fig:challengesABFG_omb}
\end{figure}

\begin{figure}[h!]
\centering
    \begin{center}
  \textbf{BBN Priors ABFG \& D \& Patchy}
    \end{center}
\includegraphics[width=0.695\textwidth,draft=false]{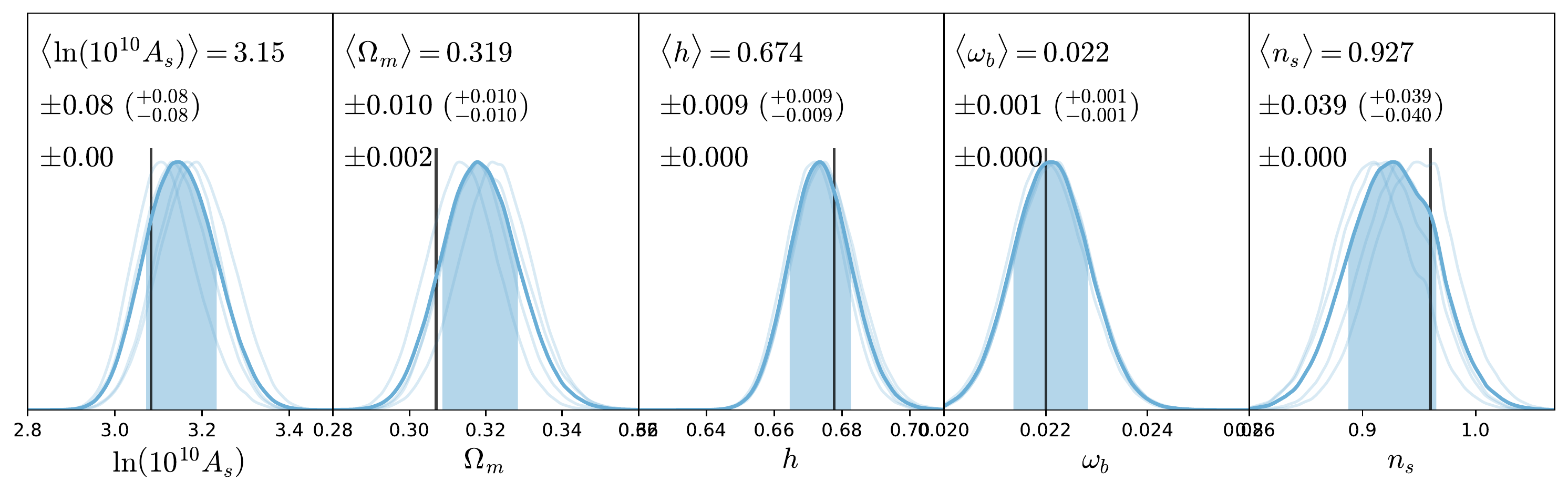}
\includegraphics[width=0.695\textwidth,draft=false]{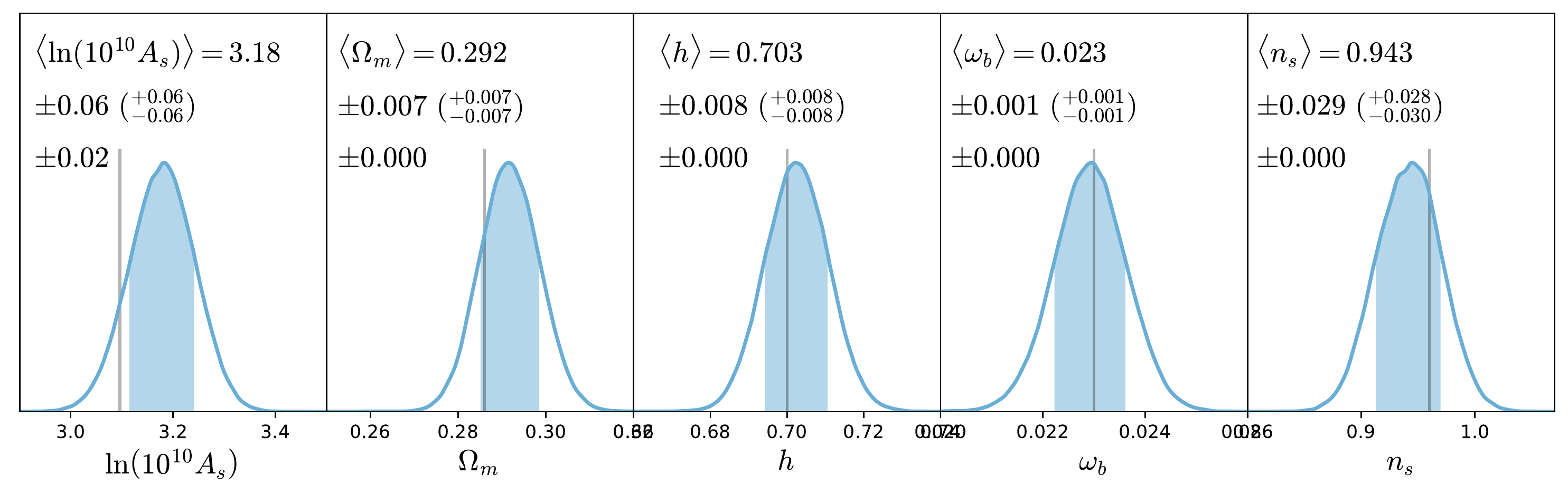}
\includegraphics[width=0.695\textwidth,draft=false]{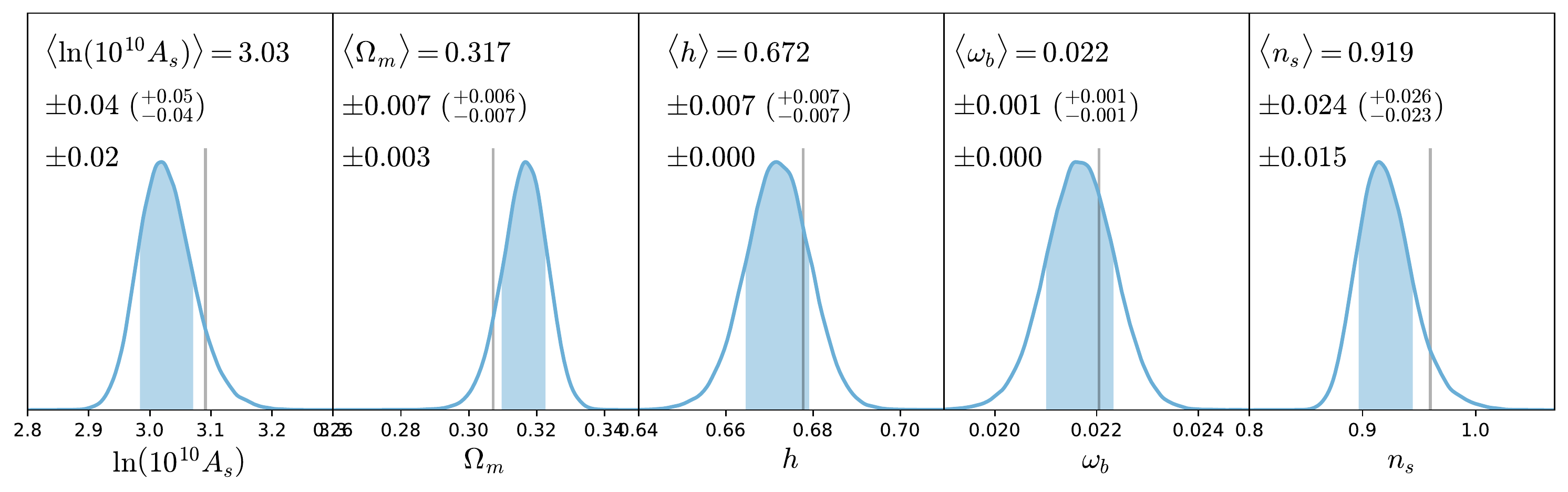}
\caption{\small Posterior distributions for the cosmological parameters obtained from the analysis of the Challenge simulations A, B, F, G ({\it top}), D ({\it middle}), and Patchy ({\it bottom}), using the Taylor expansion approximation for the EFT power spectrum, and by imposing the BBN prior on $\omega_b=\Omega_b h^2$ at $\kmax=0.23\hinvMpc$.}
\label{fig:challengeD_omb}
\end{figure}

In this section, we show the results of the analysis of the 5 Challenge simulation boxes A, B, F, G, D and of the mean of 16 Patchy mocks, which were already used in~\cite{DAmico:2019fhj}. In Fig.~\ref{fig:challengesABFG_omb}, we show the posterior distributions of the 3 cosmological parameters $\lnAs$, $\Omega_m$, $h$ from the analysis of the Challenge boxes done varying $\omega_b$, $n_s$ and the EFT-parameters with Planck2018 priors, working at $\kmax=0.2\hinvMpc$.
In Fig.~\ref{fig:challengeD_omb}, we instead show the posterior distributions of the 5 cosmological parameters $\lnAs$, $\Omega_m$, $h$, $\omega_b$ and $n_s$ from the analysis of the Challenge and Patchy boxes done varying $\omega_b$ and the EFT-parameters with a BBN prior, working at $\kmax=0.23\hinvMpc$. In this case, as we will see, the larger statistical errors allow us to work at higher  wavenumber. 

For box ABFG, if we put the Planck2018 prior on $\omega_b$ and $n_s$, we detect a small, but still negligible, systematic bias of $0.02$ on $\lnAs$ and none on $\Omega_m$ and $h$. For box D, we detect only a small systematic error on $\lnAs$ equal to 0.03, which is negligible, and none on $\Omega_m$ and $h$.
On the other hand, with a {BBN-only} prior on $\omega_b$, we have a systematic bias of $0.001$ on $\Omega_m$, which is negligible, for box ABFG, and,  for box D, a systematic error for  $\lnAs$ equal to 0.02, which is again small.

\begin{figure}[h!]
\centering
    \begin{center}
  \textbf{Patchy Planck \& BBN Priors}
    \end{center}
\includegraphics[width=0.495\textwidth,draft=false]{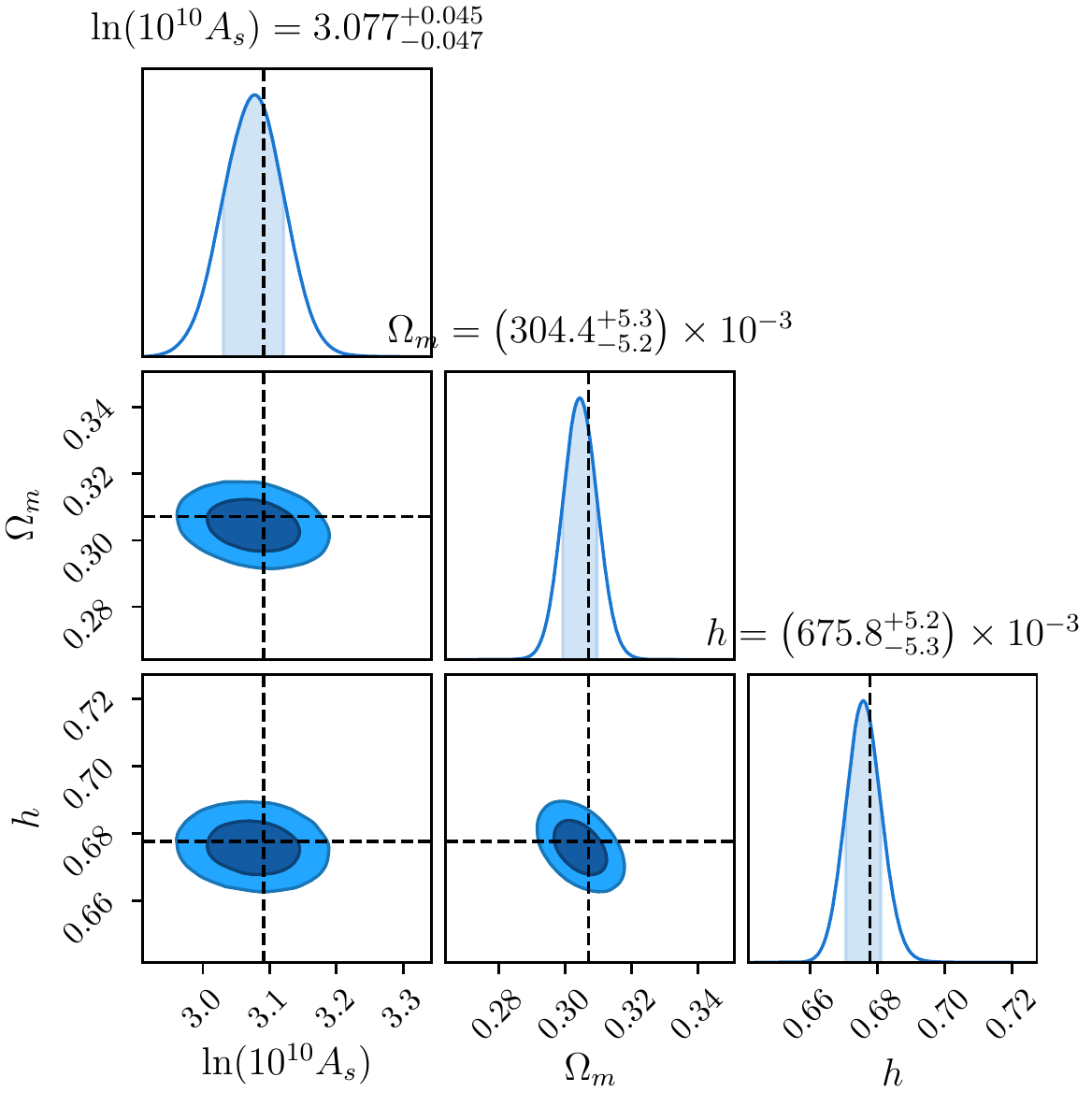}
\includegraphics[width=0.495\textwidth,draft=false]{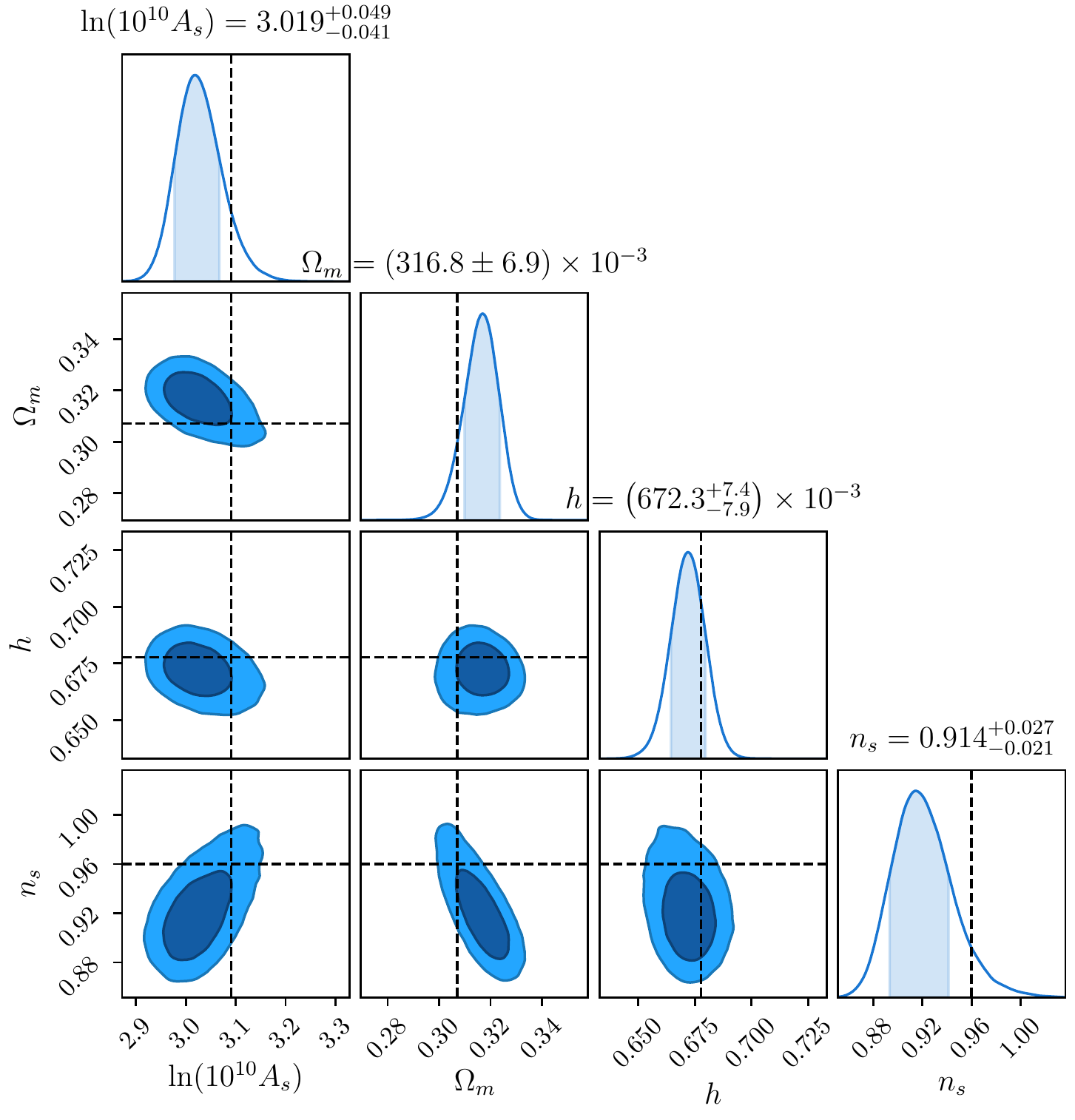}
\caption{\small Posterior distributions for the cosmological parameters obtained from the analysis of the mean of 16 Patchy mocks using the Taylor expansion approximation for the EFT power spectrum. \emph{Left:} We put the Planck2018 prior on $\omega_b=\Omega_b h^2$ and $n_s$  at $\kmax=0.20\hinvMpc$. In vertical dashed we plot the expectation value from~\cite{DAmico:2019fhj}.  \emph{Right:} We instead put the BBN prior on $\omega_b=\Omega_b h^2$ at $\kmax=0.23\hinvMpc$.}
\label{fig:patchy_omb}
\end{figure}

If we combine the systematic errors between the measurements on boxes ABFG and the ones on box D as done in~\cite{DAmico:2019fhj}, in the case of Planck priors we find only a systematic error on $\lnAs$ of 0.04, which is sufficiently negligible being $\sim \sigma_{\rm stat, \, CMASS x LOWZ}/5$ and an even more negligible systematic error on $\Omega_m$ of 0.002.
In the case of {BBN-only} priors we have a similar systematic error on $\lnAs$ of 0.03, and a systematic error on $\Omega_m$ of 0.003, which are both negligible.
In Fig.~\ref{fig:patchy_omb}, we show the analysis on the mean of 16 Patchy mock simulations, from which it is apparent that we detect no or negligible systematic error. In particular for the case of {BBN-only} prior we detect a systematic error in $n_s$ equal to $0.015\simeq \sigma_{\rm stat, \, CMASS x LOWZ}/5$, which is negligible\footnote{For the case of BBN prior with the additional bound $\sum_{i} m_{\nu_i}\leq 0.25\,{\rm eV}$, this systematic error of 0.015 is about $1/4$ of the statistical error, which is still safely negligible.}.
In conclusion, we verify that the systematic error is negligible with respect to the errors on the data at the  $\kmax$'s under consideration.

\section{Additional Results\label{app:CMASS}}

\begin{figure}[h!]
\centering
\includegraphics[width=0.495\textwidth,draft=false]{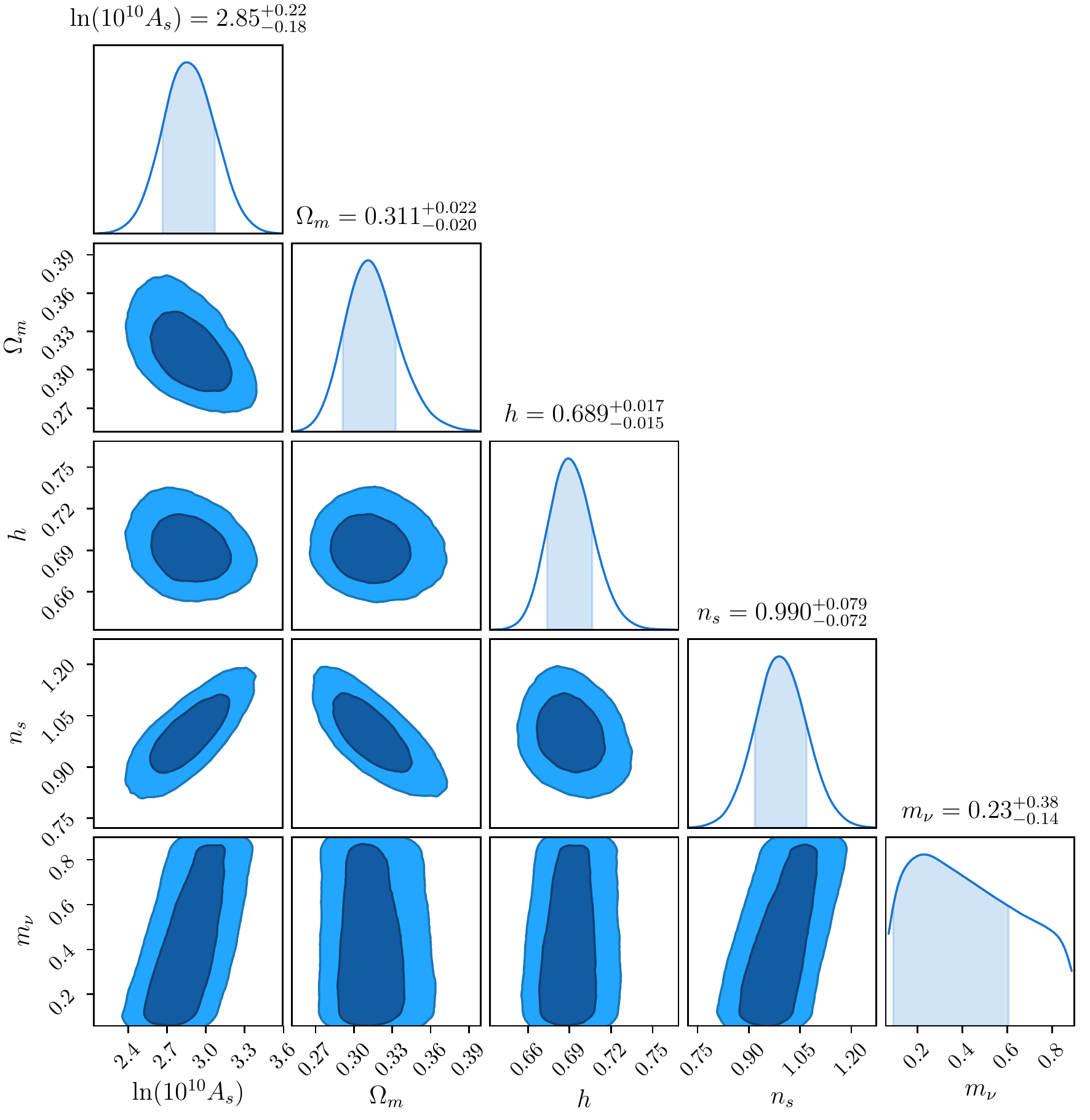}
\includegraphics[width=0.495\textwidth,draft=false]{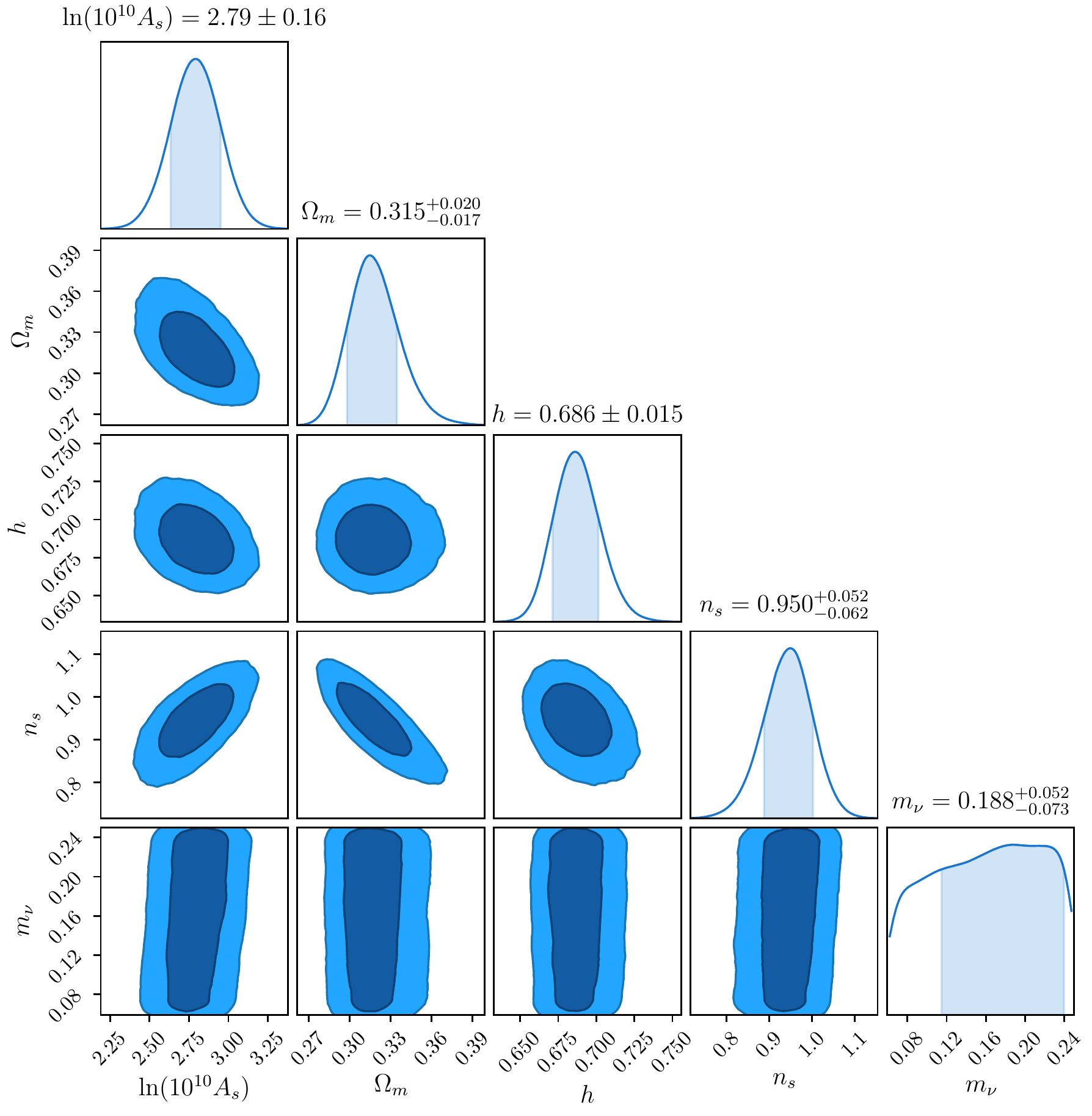}
\caption{\small \emph{Left:} Posterior distributions for the cosmological parameters obtained from the analysis of the CMASS sample, using the Taylor expansion approximation for the EFT power spectrum. We put a BBN prior on $\omega_b=\Omega_b h^2$, as explained in the text, working at $\kmax=0.23\hinvMpc$. Comparing with Fig.~\ref{fig:analysis_loose_prior}, we can see the compatibility of the datasets.
\emph{Right:} Same as on the left, but analyzing the combination CMASS$\times$LOWZ NGC ($\kmax=0.23\hinvMpc$ for CMASS and $\kmax=0.20\hinvMpc$ for LOWZ NGC) and putting an additional flat prior for the neutrino masses: $0.06\, {\rm eV}\leq \sum_{i} m_{\nu_i}\leq 0.25\,{\rm eV}$.}
\label{fig:analysis_loose_prior2}
\end{figure}

Here we briefly present some additional results. On the left of Fig.~\ref{fig:analysis_loose_prior2} we give the result of the analysis with the {BBN-only} priors on the CMASS sample, to show that it gives similar results as the CMASS$\times$LOWZ NGC analysis. This shows the compatibility of the datasets. On the right of the same figure, we analyze the CMASS$\times$LOWZ NGC sample with a BBN prior on $\omega_b$, and impose an additional flat prior for the neutrino masses: $0.06\, {\rm eV}\leq \sum_{i} m_{\nu_i}\leq 0.25\,{\rm eV}$, as inspired by Planck2018.

\bibliography{references}

\end{document}